\documentclass[letterpaper,journal]{IEEEtran}
\usepackage{amsmath,amsfonts,amssymb}
\usepackage{array}
\usepackage[caption=false,font=normalsize,labelfont=sf,textfont=sf]{subfig}
\usepackage{textcomp}
\usepackage{stfloats}
\usepackage{url}
\usepackage{verbatim}
\usepackage{graphicx}
\usepackage{color,soul}
\usepackage{lipsum}
\usepackage{mathtools}
\usepackage{mathrsfs}
\usepackage{tabularx}
\newcolumntype{Y}{>{\centering\arraybackslash}X}
\usepackage{amssymb}
\usepackage{mathtools}
\usepackage{cite}
\usepackage[colorlinks=true, linkcolor=blue, citecolor=blue, urlcolor=blue]{hyperref}
\usepackage{afterpage}

\usepackage{cuted} 
\usepackage{glossaries}
\makeglossaries
\usepackage{soul}
\usepackage{xcolor}
\usepackage{algorithm}
\usepackage{algpseudocode}
\floatname{algorithm}{Procedure}
\usepackage{enumitem}

\soulregister{\cite}{7}
\soulregister{\ref}{7}
\soulregister{\eqref}{7}
\newacronym{isp}{ISP}{internet service provider}
\newacronym{poi}{POI}{point of interest}
\newacronym{od}{OD}{origin destination}
\newacronym{rog}{ROG}{radius of gyration}

\newcolumntype{M}[1]{>{\centering\arraybackslash}m{#1}}
\def\BibTeX{{\rm B\kern-.05em{\sc i\kern-.025em b}\kern-.08em
    T\kern-.1667em\lower.7ex\hbox{E}\kern-.125emX}}
\usepackage{balance}
\begin{document}
\title{Analysis of Multi-Tone, Multi-Conductor, Spatially Discrete Traveling-Wave Modulated Loop Networks}
\author{Amirhossein Babaee, Zachary Fritts, Steve M. Young, Anthony Grbic
\thanks{This work has been supported by US Air Force grant FA8650-22-D-5406 through a subcontract with Azimuth Corporation.
Also, we are thankful for NIWC Pacific’s and IARPA’s support via contract N6600122C4507 to SRI International. The University of Michigan is a subcontractor to SRI International for the EQuAL-P program. The views and conclusions contained herein are those of the authors and should not be interpreted as necessarily representing the official policies, either expressed or implied, of ODNI, IARPA, or the U.S. Government. The U.S. Government is authorized to reproduce and distribute reprints for governmental purposes notwithstanding any copyright annotation therein. \\
The authors are with the Department of Electrical Engineering and Computer Science, University of Michigan, Ann Arbor, MI 48109 USA (e-mail: ababaee@umich.edu; zfritts@umich.edu; yms@umich.edu; agrbic@umich.edu).}}
\maketitle
\begin{abstract}
This work presents a semi-analytical framework for analyzing spatially discrete traveling-wave modulated (SDTWM) loop networks, which exhibit cavity-like behavior and support discrete spatiotemporal modes.
We introduce a computationally efficient method, based on the Interpath Relation, to analyze periodic networks using a single unit cell.
This allows characterization of driven systems with single-tone, multi-tone, and multi-conductor loop configurations.
The framework captures both multi-modal and multi-frequency harmonic interactions, and is extended to compute the spatial Green’s functions of such loop networks using analytic array scanning.
The analysis of example designs, such as an electrically small antenna and a non-magnetic circulator, is presented.
These examples confirm that the proposed approach is computationally efficient and offers physical insight, making it well-suited for the optimization of multifunctional and nonreciprocal SDTWM electromagnetic systems.
\end{abstract}

\section{Introduction}
\IEEEPARstart{S}{pace–time modulated} electromagnetic structures have attracted significant interest in the microwave and optics communities due to their ability to enable nonreciprocal responses, frequency conversion, and parametric amplification \cite{047001, 2022AdPho4a4002G, doi:10.1126/sciadv.adg7541, 202001594, 8854331, 9019736, 10041963}.

Recent works have introduced practical circuit-based realizations of space–time modulated structures through spatially discrete traveling-wave modulation (SDTWM), in which the modulation is staggered in time across an array of discrete unit cells \cite{Babaee2024, 10015153}.
This discretized traveling-wave modulation can be implemented using lumped or distributed elements.
SDTWM networks support Bloch-Floquet modes that are tailored through spatial and temporal modulation.
Their analysis can be simplified by enforcing the Interpath Relation at the unit-cell level.
The Interpath Relation is a space-time boundary condition that relates the fields at the terminals (ends) of a single unit cell of an SDTWM network.
It allows the behavior of a periodic SDTWM system to be captured by analyzing a single unit cell \cite{10015153}.
It dictates phase delays across the unit cell for both the fundamental and higher-order harmonics, enabling efficient analysis.

\begin{figure}[!t]
\centering%
\subfloat[]{%
\centering
\includegraphics[width=90mm]{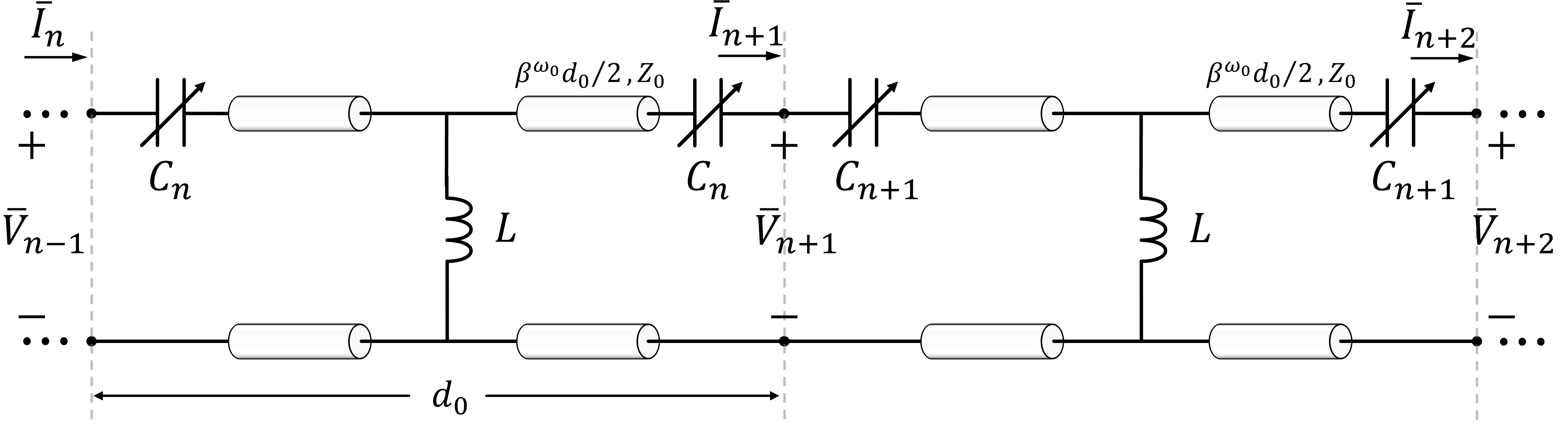}
\label{fig:1_a}
}%
\\[-0.1mm]%
\subfloat[]{%
\centering
\includegraphics[width=70mm]{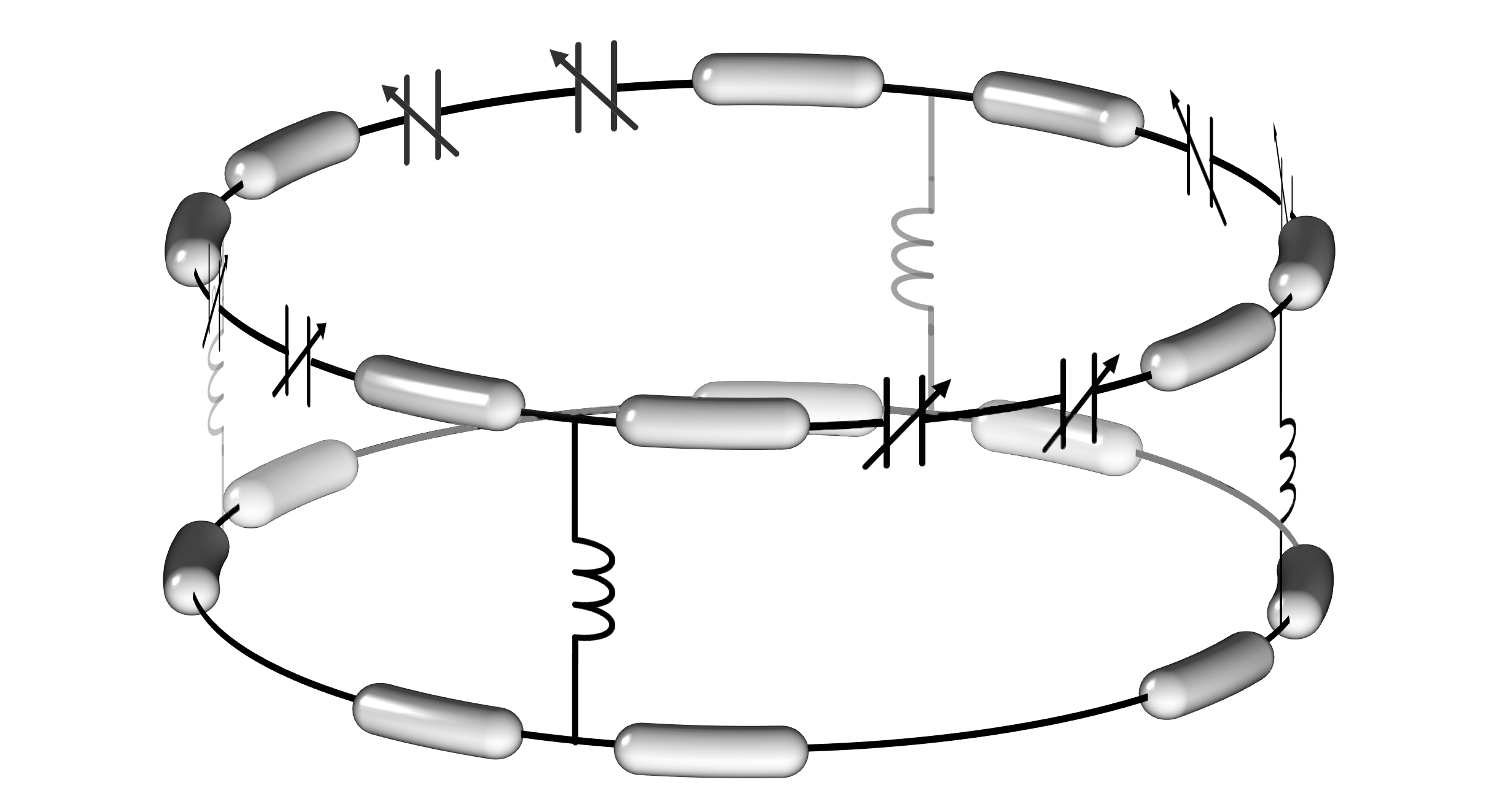}
\label{fig:1_b}
}%
\caption{A spatially discrete traveling-wave-modulated backward-wave network: (a) Straight configuration and (b) Loop configuration.}
\label{fig:1}
\end{figure}

In microwave engineering, loop configurations are widely used to implement components such as resonators, couplers, power dividers, and circulators \cite{Pozar2011, 1125612}. Conventional circulators rely on ferrite materials biased by external magnetic fields to break reciprocity. However, such designs are bulky, expensive, and incompatible with integrated circuit technologies.
To address these limitations, magnetless circulators have been developed based on space–time modulation \cite{9257419, 7378329, Sounas2013}.
By embedding modulation into loop networks, these devices achieve non-reciprocity without magnetic bias, offering a scalable and CMOS-compatible solution \cite{Cui2017Information, PhysRevApplied.16.014044}.

In this paper, we present a framework for analyzing SDTWM loop networks.
{First, we extend the driven modal solution from \cite{Babaee2024} to loop configurations, where the loop is excited by a phased array of sources.}
Section~\ref{sec:3_1} presents the driven modal solution for single-tone modulation.
{Based on this formulation, we propose a parametric amplifier design. 
We show that, by exciting a loop SDTWM network with only four unit cells, high gain can be achieved.}

{In Section~\ref{sec:3_2}, we introduce the analysis of multi-tone modulation in SDTWM loop networks.}
By applying different modulation tones or combining different spatial periods under a shared temporal modulation, the network can support multiple functions, each associated with a specific tone \cite{10785556, Wang2023, Zhu2019, PhysRevApplied.14.064060}.
{As an example, we propose a design based on the same four-unit-cell loop network. When the network is excited, the input signal is down-converted to a lower frequency and simultaneously amplified through two-tone modulation.}

{In Section~\ref{sec:3_3}, we introduce a formulation for multi-conductor SDTWM networks, enabling the excitation and control of multiple coupled modes and frequency harmonics within a structure \cite{739231, 9178345}.}
{This new contribution enables multi-modal SDTWM networks, expanding the design space to support mode-selective gain, directional coupling between conductors, and the emergence of exceptional points \cite{Baharian2023, 8007255, 9178345, 7756320, doi:10.1126/science.aae0330}.}
It also offers a circuit-based framework to model complex electromagnetic media with anisotropic or topological properties \cite{805896}.
{As an example, we consider the space–time varying electrically small antenna reported in \cite{10785556}, which can be modeled using an equivalent multi-conductor circuit representation.}

{Lastly, in Section~\ref{sec:4}, we extend the spatial Green’s function formulation using the analytic array scanning method proposed in \cite{Babaee2024} to SDTWM loop networks. 
This is achieved by combining the driven modal solutions of the loop network.}
{As a demonstration, we design a non-magnetic circulator. 
We show how the driven modal solutions and the spatial Green’s function can be used to optimize the circulator’s bandwidth and insertion loss.}

{Finally, the Appendix provides a detailed derivation of the multi-harmonic ABCD matrix used for the SDTWM unit cell.
The formulation is presented for both single-tone and multi-tone modulation and explains the construction of the multi-harmonic matrices used in the analysis.
This detailed description facilitates reproducibility of the results.}

\section{Interpath Relation in SDTWM Loop Networks} \label{sec:2}
An infinite SDTWM circuit network is illustrated in Fig.~\ref{fig:1_a}.
In this network, the modulation of the $(n+1)^{\text{th}}$ unit cell is delayed relative to the $n^{\text{th}}$ unit cell by a time interval $t_0$.  
Being linear and periodically time-varying (LPTV), the voltage and current at the input terminals of the unit cell can be related to those at the output terminals using a space-time boundary condition.  
This boundary condition, known as the Interpath Relation, enables the modeling of an infinite SDTWM network with a single unit cell \cite{10015153, Babaee2024}.
In Fig.~\ref{fig:1_a}, the capacitance of the $n^{\text{th}}$ unit cell may be any periodic-in-time function with an angular frequency $\Omega$ and progressive delay between adjacent unit cells.
The examples in this work assume a specific form for the capacitance $C_n(t) = C_0\left[1 + 2M \cos\left(\Omega t - n\Omega t_0\right)\right]$, where $C_0$ is the average capacitance, $M$ is the modulation depth, defined within the range $[0, 0.5]$, $\Omega$ is the angular modulation frequency, and $\Omega t_0$ represents the modulation phase delay per unit cell.
While the analysis presented applies to any time periodic traveling waveform, all examples in this work follow this specific form.

Assuming a fundamental angular signal frequency $\omega_0$, the modulated capacitance causes frequency mixing, so that the terminal voltages and currents can be expressed as a Fourier series,
\begin{equation}
\mathrm{v}_n(t) = \Re \left\{{\sum_{\ell={-L}}^{L} V_{n,\ell}~e^{j \left(\omega_0 + \ell \Omega \right)t}} \right\}
\label{eq:0}
\end{equation}
The total voltage and current for $n^{\text{th}}$ unit cell can be represented compactly as vector of the harmonic amplitudes $\bar{V}_n$ and $\bar{I}_n$, respectively, where $q = 2L + 1$ denotes the total number of frequency components with harmonics indexed by $\ell \in \{ -L, -L+1 \dots, L \}$.

\begin{figure}
\centering
\subfloat[]{%
\includegraphics[width=3.2in]{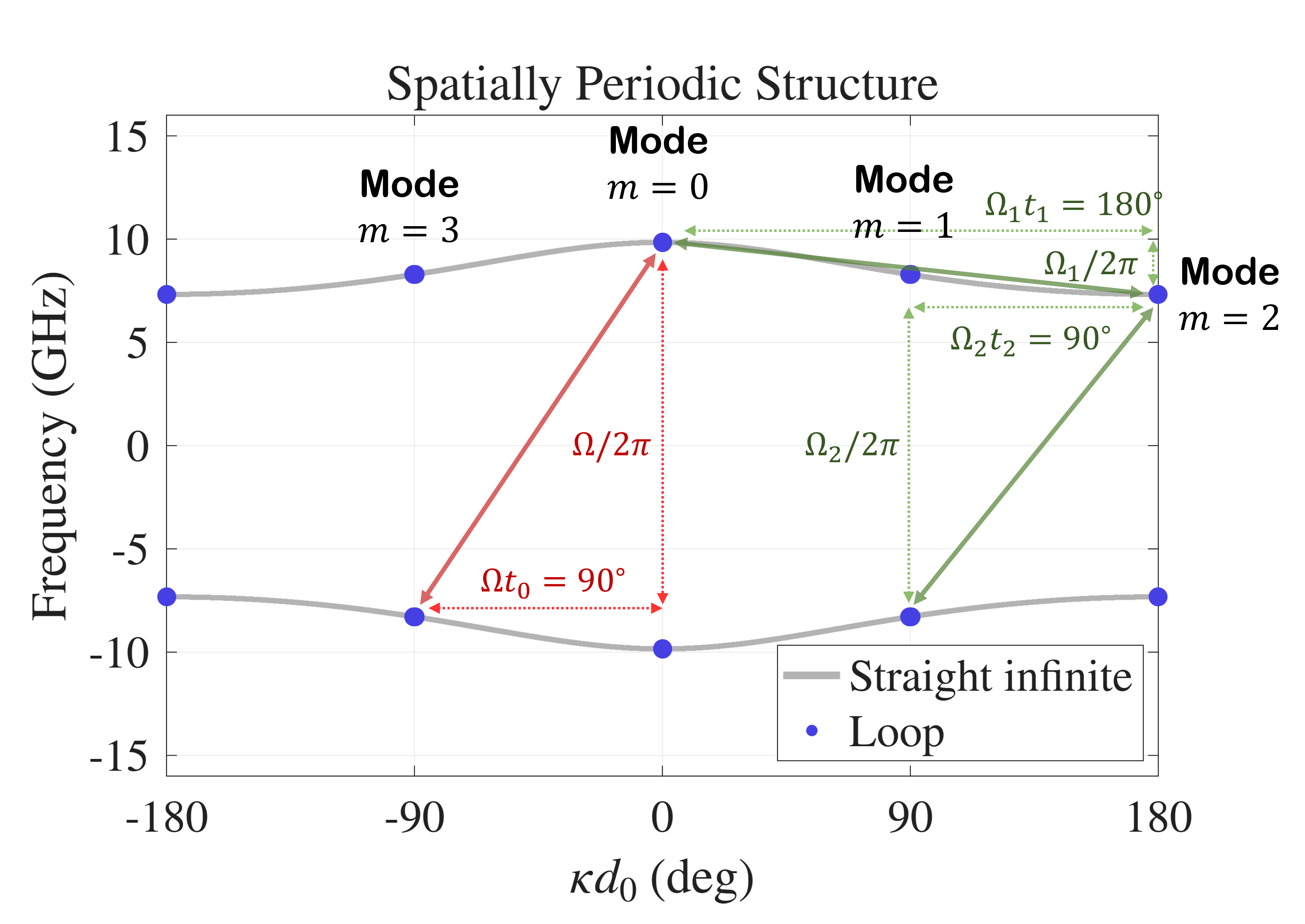}
\label{fig:2_a}
}%
\vspace{0.5em}
\subfloat[]{%
\includegraphics[width=3.0in]{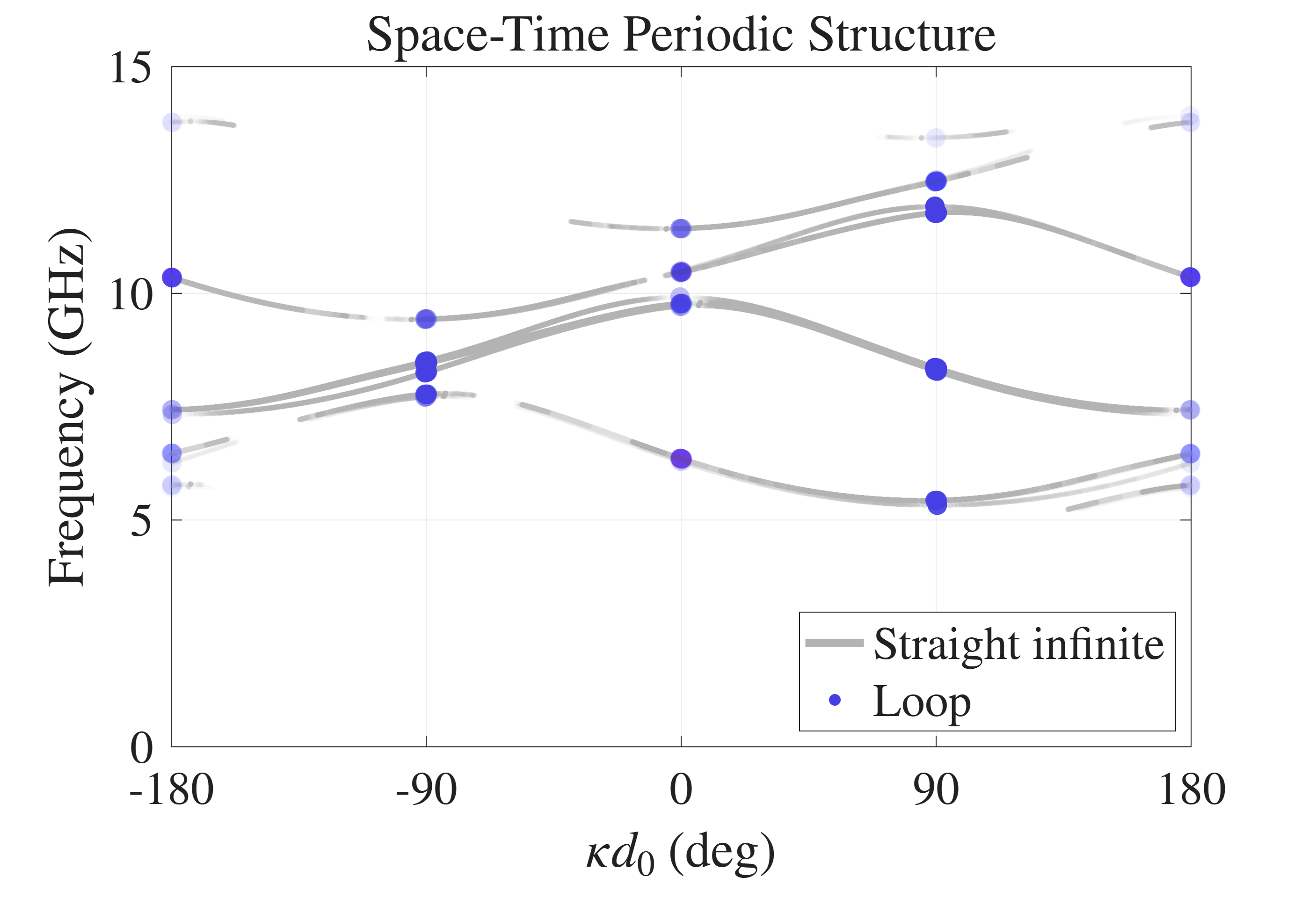}
\label{fig:2_b}
}%
\caption{{Dispersion diagrams. (a) Dispersion diagram for the static (time-invariant) network of Fig.~\ref{fig:1}. The dispersion diagram is plotted to include negative frequencies. The network has a capacitance $C_0=0.43$ pF, an inductance of $0.23$ nH, a non-dispersive transmission line electric length of $\beta^{\omega_0} d_0=32.55^\circ$ at a frequency of 10 GHz, and a characteristic impedance of $Z_0=131 \, \Omega$.
The dispersion of the infinite network of straight configuration, shown in Fig.~\ref{fig:1_a}, is plotted with the continuous gray line, while that of the loop network in Fig.~\ref{fig:1_b} is plotted with blue dots.
The arrows indicate the mode couplings used in the examples.
The red arrows correspond to the single-tone parametric amplifier example, while the green arrows correspond to the two-tone modulated parametric amplifier.
(b) Dispersion diagram of a SDTWM network with 20\% modulation depth ($M = 0.1$), a modulation frequency of 2\,GHz, and a modulation phase delay of $\Omega t_0 = 90^\circ$ between neighboring capacitors.}}
\label{fig:2}
\end{figure}

The Interpath Relation governs wave propagation within the SDTWM networks, and can be expressed as follows,
\begin{equation}
\begin{bmatrix} \bar{V}_{n+1} \\ \bar{I}_{n+1} \end{bmatrix} = 
e^{-j\kappa d_0}
\begin{bmatrix} \bar{\bar{\mathfrak{D}}} & \bar{\bar{0}} \\ \bar{\bar{0}} & \bar{\bar{\mathfrak{D}}} \end{bmatrix} 
\begin{bmatrix} \bar{V}_{n} \\ \bar{I}_{n} \end{bmatrix} =
e^{-j\kappa d_0} \bar{\bar{\mathcal{D}}}
\begin{bmatrix} \bar{V}_{n} \\ \bar{I}_{n} \end{bmatrix}
\label{eq:1}
\end{equation}
The Bloch-Floquet phase delay, $\kappa d_0$, defines the phase progression of the signal within the network.
Moreover, the delay matrix $\bar{\bar{\mathfrak{D}}}$ quantifies the added phase shifts introduced by the modulation.  
{This diagonal matrix has components defined by the Kronecker delta function as $\mathfrak{D}_{\ell,\ell'} = \delta_{\ell-\ell'} e^{-j\ell\Omega t_0}$, where $\ell$ denotes the frequency harmonic index and $\delta_{x}$ is the Kronecker delta function, i.e., $\delta_{0}=1$ and $\delta_{x}=0$ for $x\neq0$.}
The delay matrix accounts for the modulation phase delay at the $\ell^{\,\text{th}}$ frequency harmonic, given by $\ell\Omega t_0$.

Now, consider forming a loop by connecting the input terminal of the zeroth unit cell to the output terminal of the $N$-th unit cell, as shown in Fig.~\ref{fig:1_b}, with $N = 4$.
The loop forms a cavity that imposes a discrete set of allowable phase delays determined by the total number of unit cells in the network ($N$).  
As voltages and currents progress through the loop network in accordance with the Interpath Relation (\ref{eq:1}) and return to their starting point, they must satisfy an azimuthally periodic condition, also known as Born–von Karman boundary condition \cite{Ashcroft76}.
This condition requires the Bloch-Floquet modes ($\kappa d_0$) and the modulation phase delay ($\Omega t_0$) to be an integer multiple of $2\pi/N$.
Thus, the supported Bloch-Floquet modes and modulation phase delays are given by
\begin{equation}
\begin{aligned}
&\kappa d_0 = m \frac{2\pi}{N} \qquad \qquad \quad &{m=0,~1,~\cdots,~N-1}&
\\
&\Omega t_0 = p \frac{2\pi}{N} \qquad \qquad \quad & 
{p=0,~1,~\cdots,~N-1}&, 
\label{eq:2}
\end{aligned}
\end{equation}
where $m$ is the mode number and $p$ is the modulation mode number.

Fig.~\ref{fig:2} shows the dispersion diagram for the backward-wave network depicted in Fig.~\ref{fig:1_b} \cite{1444496, 1447705}.
The solid traces represent the dispersion of the corresponding straight infinite network (Fig.~\ref{fig:1_a}), indicating continuous dispersion, while the dots mark discrete modes that satisfy the loop boundary condition given by \eqref{eq:2}.
Since in Fig.~\ref{fig:2} the loop network consists of four unit cells ($N = 4$), the allowed values of $\kappa d_0$ are $\{0, \frac{\pi}{2}, \pi, -\frac{\pi}{2}\}$, corresponding to $m = 0, 1, 2, 3$, respectively.
The modes have distinct resonant frequencies, except for $m = 1$ and $m = 3$, which are degenerate.
The space–time modulated dispersion in Fig.~\ref{fig:2_b} is plotted using the energy-weighted SDTWM method introduced in \cite{Babaee2024}.
The energy-weighted method uses the energy computed from the modal voltages and currents, found from the eigenvectors.
It emphasizes the most significant harmonic components of each Bloch-Floquet mode.
The contrast used to plot the dispersion diagram is proportional to the energy of the mode. 

To summarize, forming a loop from an SDTWM network imposes discrete phase delays, restricting the supported Bloch-Floquet modes to a discrete set.

\section{Driven Modal Solutions of SDTWM Loops} \label{sec:3}
This section analyzes SDTWM loop networks excited by a phased array of sources. In other words, each unit cell contains a current source with a progressive phase delay, $\varphi_s$.
This analysis follows that developed in~\cite{Babaee2024} for straight SDTWM networks, and is extended to address driven modal solutions under multi-tone modulation and to multi-conductor SDTWM loops.

\begin{figure}[!t]
\centering%
\subfloat[]{%
\centering
\includegraphics[width=50mm]{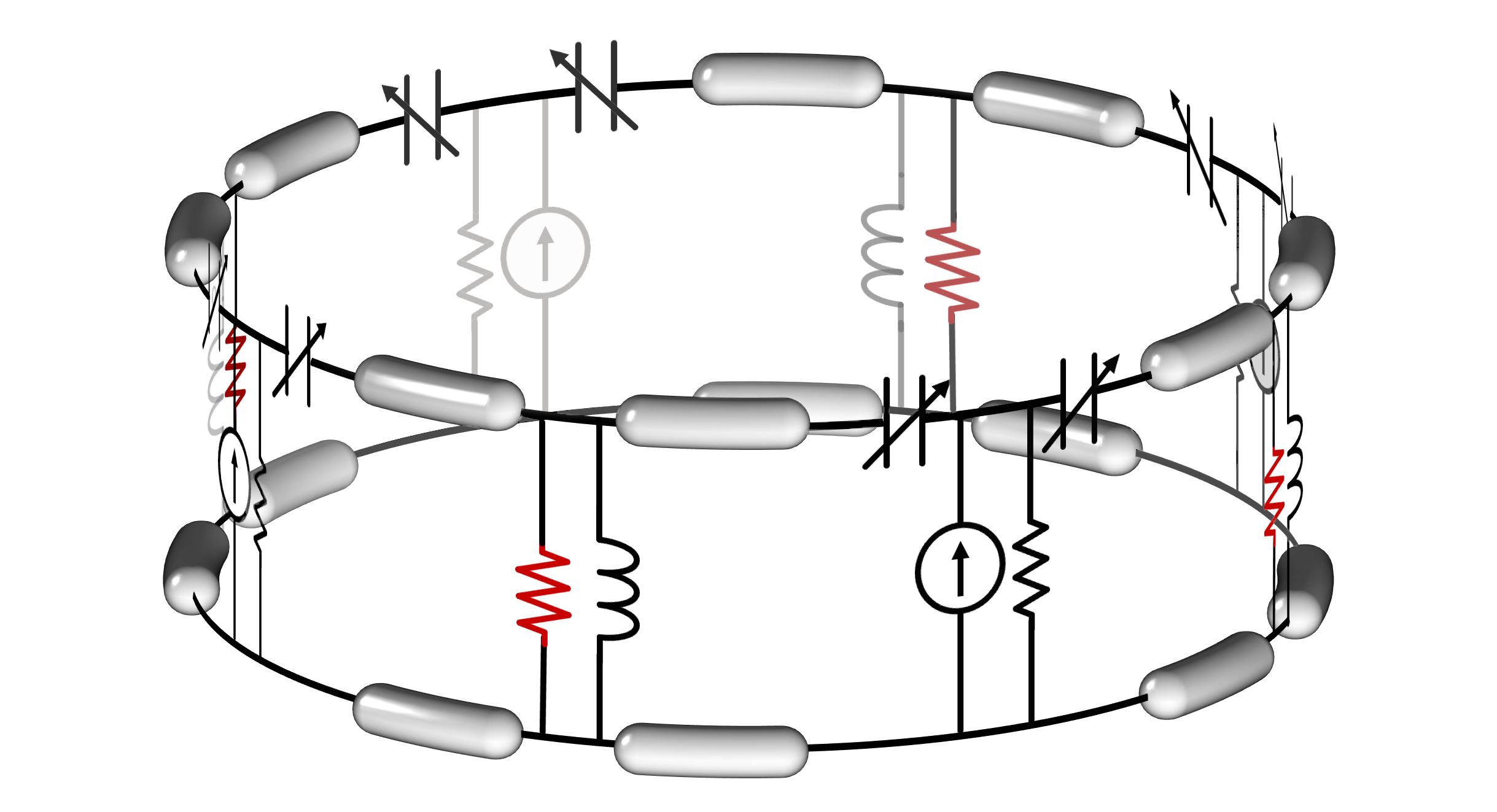}
\label{fig:3_a}
}%
\\[-0.1mm]%
\subfloat[]{%
\centering
\includegraphics[width=80mm]{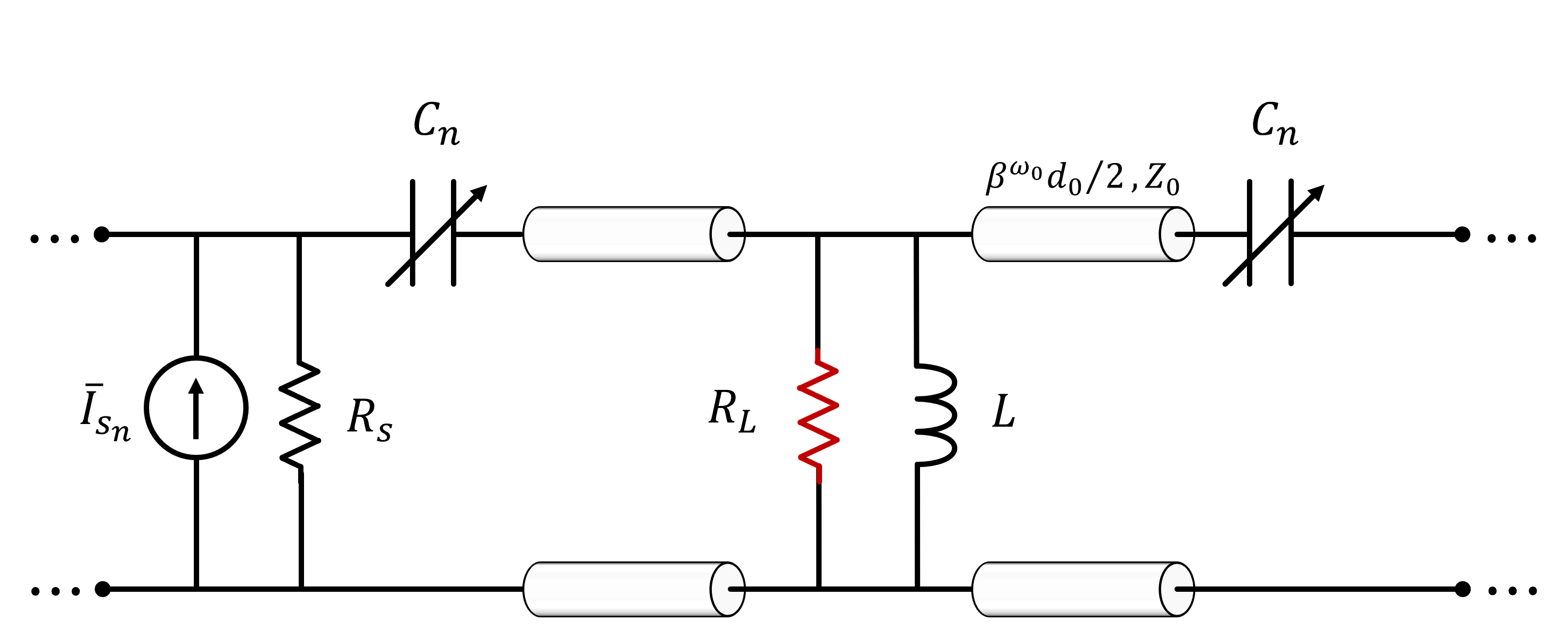}
\label{fig:3_b}
}%
\caption{SDTWM loop network excited by a phased array of sources.  
(a) A SDTWM loop network excited by a phased array of current sources, each modeled as a Norton equivalent with a source resistor $R_s$.  
(b) The $n^{\text{th}}$ unit cell of the SDTWM loop network, where the input is a current source with source resistor $R_s$, and the output is the resistive load $R_L$, shown in red.
}
\label{fig:3}
\end{figure}

Fig.~\ref{fig:3_a} illustrates a backward-wave SDTWM loop network excited by a phased array of current sources.  The $n^{\text{th}}$ unit cell of the loop is shown in Fig.~\ref{fig:3_b}.
A shunt resistor $R_L$ (in red) is included to serve as the output load in the following examples.
We analyze the case where the source current in the $n^{\text{th}}$ unit cell is a periodic function of the form
\begin{equation}
i_{s_n}(t) = \Re \left\{ \sum_{\ell = -L}^{L} I_{s_{n,\ell}}~ e^{j \left(\omega_0 + \ell \Omega \right)t - jn\varphi_s} \right\},
\label{eq:Is}
\end{equation}
where $I_{s_{n,\ell}}$ is the current amplitude of the $\ell^{\,\text{th}}$ harmonic at the $n^{\text{th}}$ unit cell, and $\varphi_s$ denotes the progressive phase shift between adjacent unit cells.

Using a multi-harmonic vector representation, the current source at the $n^{\text{th}}$ unit cell can be compactly expressed as
\begin{equation}
\bar{I}_{s_n} = \bar{I}_{s_0} e^{-j n \varphi_s},
\label{eq:3}
\end{equation}
where $\bar{I}_{s_0}$ denotes the source amplitude across all frequency harmonics.
Each unit cell is modeled using a Norton source (a current source in parallel with a resistor) with a progressive phase shift ($\varphi_s$).
For simplicity in \eqref{eq:Is}, we assume the same progressive phase shift ($\varphi_s$) for all excited frequency harmonics.
However, the progressive phase shift can be generalized such that each harmonic undergoes a distinct phase delay.
Note that the approach is not dependent on representing embedded sources using a Norton equivalent. 
Any source configuration that maintains a consistent phase progression across the cells can be accommodated.

In contrast to the source-free case discussed in Sec.~\ref{sec:2}, where the Bloch-Floquet modes ($\kappa d_0$) represent the natural eigenmodes of the SDTWM loop network, the phased-array excitation introduces an impressed phase delay, $\varphi_s$, across the unit cells.
The excitation corresponds to a mode with an azimuthal order that satisfies the loop boundary condition defined in \eqref{eq:2}.
Specifically, the allowed source phase delay is  quantized as,
\begin{equation}
\varphi_s = m \frac{2\pi}{N}, \qquad \quad m = 0,1,\cdots,~N-1,
\label{eq:4}
\end{equation}
where $m$ denotes the azimuthal order of the source excitation.

Since the phased array of sources is periodic, the Interpath Relation still governs the network's response.  
As a result, the Interpath Relation in~\eqref{eq:1} can be written for a unit cell in the driven case as follows,
\begin{equation}
\begin{bmatrix} \bar{V}_{n+1} \\ \bar{I}_{n+1} \end{bmatrix}
=
e^{-j\varphi_s} \bar{\bar{\mathcal{D}}}
\begin{bmatrix} \bar{V}_{n} \\ \bar{I}_{n} \end{bmatrix}
=
e^{-jm \frac{2\pi}{N}} \bar{\bar{\mathcal{D}}}
\begin{bmatrix} \bar{V}_{n} \\ \bar{I}_{n} \end{bmatrix}
\label{eq:5}
\end{equation}
Here $\varphi_s$ is a phase associated with the driving sources, while in \eqref{eq:1} the corresponding phase was a modal phase.
\subsection{Single-Tone Modulation SDTWM Network} \label{sec:3_1}

First, let us consider single-tone modulation, where the parametric capacitance of the $n^{\text{th}}$ unit cell is defined as $C_n(t) = C_0 \left[1 + 2M \cos\left(\Omega t - n\Omega t_0\right)\right]$.
As described earlier, the cosine waveform serves as a specific example.
However, any periodic waveform in time with a progressive phase shift in space can be analyzed using this framework.

The driven modal solution for an infinite SDTWM network of straight configuration under such modulation was presented in~\cite{Babaee2024}.
For the loop configuration considered here, the analysis proceeds similarly.
Using the ABCD matrix of the $n^{\text{th}}$ unit cell, which relates the input and output voltages and currents, and applying Kirchhoff's Current Law (KCL) and Kirchhoff's Voltage Law (KVL) at the input, we can write
\begin{equation}
\begin{bmatrix} \bar{V}_{n} \\ \bar{I}_{n} \end{bmatrix} +
\begin{bmatrix} \bar{0} \\ \bar{I}_{s_n} \end{bmatrix} =
\bar{\bar{T}}_{n}
\begin{bmatrix} \bar{V}_{n+1} \\ \bar{I}_{n+1} \end{bmatrix}.
\label{eq:ABCD}
\end{equation}
where $\bar{\bar{T}}_{n}$ is the ABCD matrix of the $n^{\text{th}}$ unit cell.

Using the Interpath Relation in \eqref{eq:5}, the output voltages and currents can be expressed in terms of the input ones.
By substituting \eqref{eq:5} into the right-hand side of \eqref{eq:ABCD}, we obtain
\begin{equation}
\begin{bmatrix} \bar{V}_{n} \\ \bar{I}_{n} \end{bmatrix} +
\begin{bmatrix} \bar{0} \\ \bar{I}_{s_n} \end{bmatrix} =
e^{-j m \frac{2\pi}{N}} \, \bar{\bar{T}}_{n} \bar{\bar{\mathcal{D}}}
\begin{bmatrix} \bar{V}_{n} \\ \bar{I}_{n} \end{bmatrix}.
\label{eq:ADBCphis}
\end{equation}

\begin{figure}[h]
\centering
\subfloat[]{%
\includegraphics[height=1.5in]{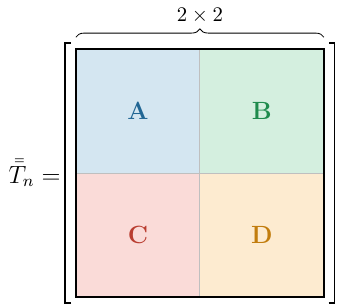}
\label{fig:Mat_1a}
}%
\vfill
\subfloat[]{%
\includegraphics[height=1.5in]{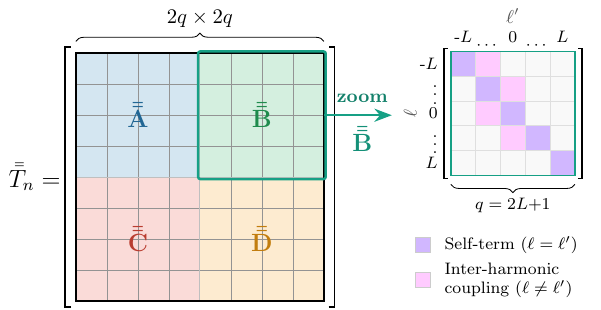}
\label{fig:Mat_1b}
}%
\caption{{ABCD matrices for (a) an LTI network and (b) a single-tone SDTWM network.}}
\label{fig:Mat_1}
\end{figure}

Given the phased array of current sources shown in Fig.~\ref{fig:3_b}, the resulting multi-harmonic voltages and currents at the input of each unit cell can then be calculated as follows \cite{Babaee2024},
\begin{equation}
\begin{bmatrix} \bar{V}_{n} \\ \bar{I}_{n} \end{bmatrix}_m =
\left( e^{-j m \frac{2\pi}{N}} \bar{\bar{T}}_{n} \bar{\bar{\mathcal{D}}}-\bar{\bar{I}} \right)^{-1}
\begin{bmatrix} \bar{0} \\ \bar{I}_{s_n} \end{bmatrix},
\label{eq:6}
\end{equation}
{In the LTI network, the ABCD matrix is a $2 \times 2$ matrix. 
Each element, for example $\mathbf{B}$, is a scalar, as shown in Fig.~\ref{fig:Mat_1a}.
In the SDTWM network, the ABCD matrix becomes a block matrix, as shown in Fig.~\ref{fig:Mat_1b}. 
Each sub-block, for example $\bar{\bar{\mathbf{B}}}$, is now a matrix that accounts for multiple harmonics.
If $q$ harmonics are included in the analysis, then $\bar{\bar{\mathbf{B}}} \in \mathbb{C}^{q \times q}$.
Each element $\bar{\bar{\mathbf{B}}}[\ell,\ell']$ is a scalar.
In each sub-block, the diagonal elements correspond to harmonic self-terms. 
The off-diagonal elements correspond to the inter-harmonic coupling.
As a result, for single-tone modulation, the ABCD matrix satisfies $\bar{\bar{T}}_n \in \mathbb{C}^{2q \times 2q}$.
The calculation of the multi-harmonic ABCD matrix for a single-tone SDTWM network is provided in the Appendix.}

Using (\ref{eq:6}), the voltages and currents at the input of all $N$ unit cells can be determined for an excitation with azimuthal order $m$.
As a result, the phased array excitation of the SDTWM loop network (see Fig.~\ref{fig:3}) can be determined using (\ref{eq:6}), referred to simply as the driven modal solution.
Leveraging the Interpath Relation, the driven modal solution requires the analysis of only a single unit cell.

\subsubsection{Example 1: Single-Tone Modulated Parametric Amplifier} \label{sec:3_1_1}

As an example, the network depicted in Fig.~\ref{fig:3}, consisting of four unit cells, is designed as a parametric amplifier.
This network is excited by a phased array of sources, and terminated (output) with shunt resistors $R_L$ shown in red (see Fig.~\ref{fig:3}).
Here, both the source resistor ($R_S$) and the load ($R_L$) are set to $50~\Omega$.

{In this network, four modes are supported.
Because the loop network behaves as a cavity resonator, these modes correspond to four resonant tanks.
In this example, parametric amplification is achieved by coupling the $m=0$ mode to the $m=3$ mode at a negative frequency.
Here, the $m=0$ mode serves as the signal (primary resonant tank), while the $m=3$ mode serves as the idler (secondary resonant tank) \cite{collin2007foundations}.
Time variation couples these two resonators, which operate at different resonant frequencies, and enables parametric amplification.}

\begin{figure}[!t]
\centering
\includegraphics[width=80mm]{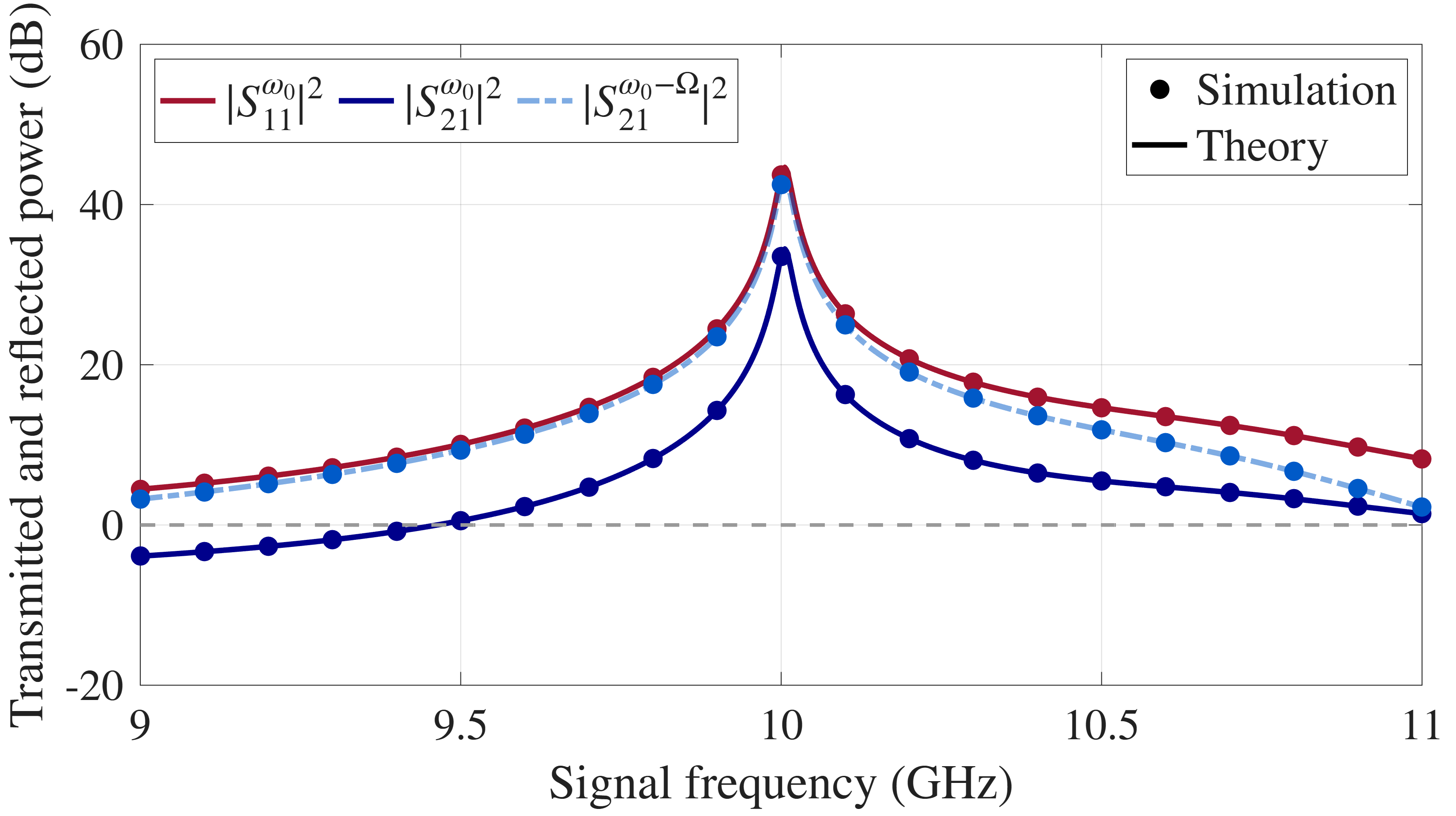}
\caption{{Transmitted and reflected power responses of a single-tone-modulated SDTWM loop network operating as a parametric amplifier. The dark and light blue lines correspond to the transmission at the signal frequency ($\omega_0$) for $m=0$ and the idler frequency ($\omega_0-\Omega$) for $m=3$, respectively. Dotted lines correspond to results from harmonic balance simulations.}}
\label{fig:4}
\end{figure}

{Once the signal and idler modes are identified, an arrow can be drawn between these modes in the lossless LTI dispersion diagram (red arrow in Fig.~\ref{fig:2_a}).
From this arrow, the required modulation parameters can be directly obtained.
To operate at the $m=0$ mode, the phased array of sources is excited at the zeroth azimuthal order, corresponding to $\varphi_s = 0^\circ$, and at the angular frequency $\omega_0$.
The modulation frequency and phase delay are selected to satisfy the phase-matching condition between the two modes.
To couple the $m=0$ mode to the $m=3$ mode at negative frequency, the required modulation frequency ($\Omega/2\pi$) is obtained from the vertical height of the arrow in Fig.~\ref{fig:2_a}.
This corresponds to the sum of the resonant frequencies of the two modes.
The required modulation phase delay is determined by the horizontal width of this arrow in Fig.~\ref{fig:2_a}, which is $\Omega t_0 = (0 - 3) \times \frac{2\pi}{4} = -\frac{3\pi}{2} = \frac{\pi}{2}$.
The circuit parameters are listed in Table~\ref{tab:1}.}

{To provide a concise and generalizable design guideline, the procedure based on the LTI dispersion diagram is summarized as follows:}
\begin{algorithm}
\caption{{LTI Dispersion-diagram-based selection of modulation parameters}}
\label{alg:dispersion_design}
\begin{algorithmic}[1]
\Require {Lossless LTI dispersion diagram}

\State {Select the desired signal mode. This determines the azimuthal order $m_s$, the signal frequency $\omega_0/2\pi$, and the required source phase progression.}

\Statex \begin{center}
{$ \varphi_s = m_s \frac{2\pi}{N} $}
\end{center}

\State {Select the idler mode to which the signal is to couple. This determines the idler mode number $m_i$ and the idler frequency $\omega_i/2\pi$, which may lie at a negative frequency.}

\State {Draw an arrow from the signal mode to the idler mode on the LTI dispersion diagram.}

\State {Choose the modulation frequency from the vertical separation between the two modes.}

\Statex \begin{center}
{$ \Omega = \left| \omega_0 - \omega_i \right| $}
\end{center}

\State {Choose the modulation phase delay per unit cell from the horizontal separation between the two modes.}

\Statex \begin{center}
{$ \Omega t_0 = \bigl(m_s - m_i\bigr)\frac{2\pi}{N} $}
\end{center}

\end{algorithmic}
\end{algorithm}

Fig~\ref{fig:4} shows the response of the network.
In this example, with only four unit cells, a high gain of approximately 40\,dB is achieved.
This example demonstrates the advantage of SDTWM loop networks over straight periodic ones, as they can achieve high gain amplification using only a few unit cells.
However, there are two main drawbacks associated with this design.
The first is the presence of a strong idler signal at the output, which is not always desirable in practice.
Second, there is significant reflection at the input.
These two effects are expected consequences of the parametric amplification process, which simultaneously amplifies both the signal and idler frequencies and introduces a large negative resistance at the input port.

The computed results are in excellent agreement with harmonic balance simulation using the commercial software ADS Keysight, shown with dots in Fig.~\ref{fig:4}.
Both the simulation and semi-analytical model use $q = 11$ harmonics. 
The implementation of time-varying capacitances in Keysight ADS follows the approach described in the supplementary material of \cite{Sounas2013}.
Although this example is not impedance matched at the source and therefore not fully optimized, it serves as a clear demonstration that the proposed analysis is accurate and valid.

\begin{table}[ht]
\centering
\caption{Design parameters for a single-tone modulated SDTWM parametric amplifier}
\label{tab:1}
\vspace{0.5em}
\begin{tabularx}{0.44\textwidth}{|Y|Y|Y|Y|Y|Y|}
\hline
$C_0$ & $L$ & $\beta^{\omega_0} d_0$ & $\omega_0/2\pi$ & $Z_0$ & $\varphi_s$ \\
\hline
0.43\,pF & 0.23\,nH & $32.55^\circ$ & 10\,GHz & 131\,$\Omega$ & $0^\circ$\\
\hline
\end{tabularx}

\vspace{0.6em}

\begin{tabularx}{0.3\textwidth}{|Y|Y|Y|}
\hline
$\Omega/2\pi$ & $\Omega t_0$ & $M$ \\
\hline
18.94\,GHz & $90^\circ$ & 0.395 \\
\hline
\end{tabularx}
\end{table}

{To summarize, this example highlights the advantage of SDTWM loop networks over straight periodic networks, since large parametric gain can be obtained using only a few unit cells.
However, two limitations are observed.
First, an amplified idler frequency appears at the output.
Second, significant reflection occurs at the input at the signal frequency.}
In the following sections, we present alternative techniques and network designs based on multi-tone modulation and multi-conductor SDTWM networks that overcome these limitations.

\subsection{Multi-Tone Modulation SDTWM Network} \label{sec:3_2}
In Section~\ref{sec:3_1}, we showed that with single-tone modulation, the signal mode can couple to only one idler mode.
This results in a network with a single functionality, which in Section~\ref{sec:3_1_1} was parametric amplification.
However, the proposed framework supports all forms of linear periodically time-varying (LPTV) modulation.
Therefore, it can be extended to handle a superposition of multiple LPTV modulations with different modulation frequencies.
The modulation can couple two distinct modes or couple a single mode to multiple idler modes.
This enables the design of multifunctional networks.

As an example, consider a multi-tone capacitance modulation consisting of the summation of $v$ cosine-modulated tones:
\begin{equation}
C_n(t) = C_0 \left[ 1 + \sum_{\nu=1}^{v} 2M_\nu \cos(\Omega_\nu t - n \Omega_\nu t_\nu) \right],
\label{eq:7}
\end{equation}
where $M_\nu$, $\Omega_\nu$, and $\Omega_\nu t_\nu$ denote the modulation depth, angular frequency, and spatial phase delay of the $\nu^{\text{th}}$ tone, respectively.
It is important to note that the cumulative modulation depths across all tones should remain below $0.5$.
This constraint ensures that the time-varying capacitance in \eqref{eq:7} does not become negative.
In practice, this modulation is typically realized using varactor diodes, switches, or similar components.
These elements cannot support negative capacitance values or forward-bias conditions.

By considering $q$ harmonics associated with each modulation tone, the total number of distinct frequency harmonics becomes $q^v$, where $v$ is the number of modulation tones.
As a result, both the transmission and delay matrices in the Interpath Relation of \eqref{eq:1} span $v$ dimensions, where each dimension corresponds to one of the modulation frequencies.

For example, the delay matrix is no longer a simple diagonal matrix of the form $\mathfrak{D}_{\ell,\ell'} = \delta_{\ell-\ell'} e^{-j \ell \Omega t_0}$.
Rather, it becomes a multi-dimensional matrix of size $q \times q \times \cdots \times q$ (with $v$ axes), resulting in a total of $q^v$ entries.
In other words, the delay matrix lies in $\mathfrak{D} \in \mathbb{C}^{q \times q \times \cdots \times q}$ ($v$ times).

However, the Interpath Relation in \eqref{eq:1} is formulated using conventional two-dimensional matrix algebra.
To enable multi-tone analysis, the $v$-dimensional delay and ABCD matrices must be appropriately mapped into equivalent two-dimensional forms.
This mapping process, which compresses the multi-index harmonic structure into linear indices, is detailed in the Appendix.
This mapping procedure can be applied to both straight infinite networks and loop networks.

\begin{figure}[t]
\centering
\subfloat[]{%
\includegraphics[height=1.7in]{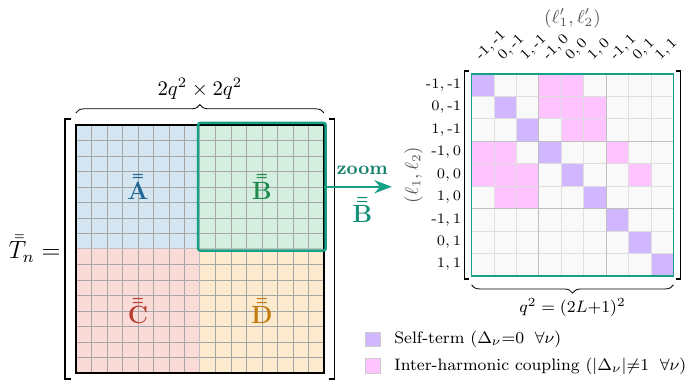}
}%
\caption{{ABCD matrix for a two-tone SDTWM network ($v=2$ and $q=3$).}}
\label{fig:Mat_2}
\end{figure}

After applying this mapping, the $\bar{\bar{\mathcal{D}}}$ matrix becomes $\bar{\bar{\mathcal{D}}} \in \mathbb{C}^{2q^v \times 2q^v}$, and the same driven modal solution described in \eqref{eq:6} can be extended directly to the multi-tone case.
The only difference is the increased size of the two-dimensional matrices and vectors involved.
Therefore, the proposed analysis framework remains structurally simple, even when generalized to multi-tone modulation scenarios.

{In Fig.~\ref{fig:Mat_2}, for a two-tone modulation ($v=2$) SDTWM network, the multi-harmonic ABCD matrix remains $2 \times 2$ block matrix, similar to the single-tone case in Fig.~\ref{fig:Mat_1b}. 
However, the sub-blocks, such as $\bar{\bar{\mathbf{B}}}$, become larger.
This is because all combinations of harmonics from the two modulation tones are included. 
In Fig.~\ref{fig:Mat_2}, $\ell_1$ and $\ell_2$ denote the harmonic indices corresponding to the first and second modulation tones. 
If $q=3$ harmonics are considered for each tone, then the total number of harmonic combinations is $3^2$.
As a result, $\bar{\bar{\mathbf{B}}} \in \mathbb{C}^{3^2 \times 3^2}$.
The construction of the multi-tone ABCD matrix is described in the Appendix.}

{One benefit of the multi-tone formulation is its ability to account for non-ideal modulation waveforms.
In practice, the modulation signal may deviate from an ideal sinusoid due to imperfections in the bias circuitry.
If the distorted waveform remains periodic, its additional harmonic components can be incorporated into the multi-tone modulation model and analyzed within the proposed framework.
In contrast, non-periodic disturbances such as noise or phase fluctuations cannot be directly represented by the harmonic expansion used in this work.}
Now, with the ability to include multi-tone modulation, SDTWM loop networks can be designed that support multiple functions simultaneously.

\subsubsection{Example 2: Two-Tone Modulated Parametric Amplifier} \label{sec:3_2_1}
{As an example, the same loop network shown in Fig.~\ref{fig:3} with the circuit parameters shown in Table.~\ref{tab:1} is redesigned using two-tone modulation.}
Here, the time-varying capacitance consists of a superposition of two cosine waveforms, as described by \eqref{eq:7}.

{Here, after selecting the signal and the idler modes, we follow the same procedure described in {Procedure~\ref{alg:dispersion_design}} to determine the required modulation parameters.
Each modulation tone couples two specific modes, as shown by the green arrows in Fig.~\ref{fig:2_a}.}

{First, we excite the zeroth azimuthal order ($\varphi_s = 0^\circ$) at angular frequency $\omega_0$.
This sets the network to operate at the $m=0$ mode.
The first modulation tone is designed to couple the $m=0$ mode (signal) to the $m=2$ mode (first idler) at a positive lower frequency.
From the lossless LTI dispersion diagram in Fig.~\ref{fig:2_a}, coupling $m=0$ to $m=2$ requires the modulation frequency to equal the difference in resonant frequencies of these two modes.
For phase matching, the required modulation phase delay is $\Omega_1 t_1 = (0 - 2) \times \frac{2\pi}{4} = -\pi = \pi$.
This corresponds to the horizontal separation between the two modes in the dispersion diagram.
This first coupling enables frequency conversion.
An input at signal ($\omega_0$) is down-converted to the first idler ($\omega_0 - \Omega_1$).}

{Next, the second modulation tone couples the $m=2$ mode (first idler) to the $m=1$ mode (second idler) at a negative frequency.
From Fig.~\ref{fig:2_a}, the required modulation frequency equals the sum of the resonant frequencies of the two modes.
For phase matching, the required modulation phase delay is $\Omega_2 t_2 = (2 - 1) \times \frac{2\pi}{4} = \frac{\pi}{2}$.
This second coupling produces parametric amplification between the first idler ($\omega_0 - \Omega_1$) and the second idler ($\omega_0 - \Omega_1 - \Omega_2$). 
Consequently, when the network is excited at angular frequency $\omega_0$, gain is expected at $\omega_0 - \Omega_1$.
The modulation parameters are summarized in Table~\ref{tab:2}, with $R_S$ and $R_L$ each assumed to be $50~\Omega$.}

\begin{figure}[!t]
\centering
\includegraphics[width=80mm]{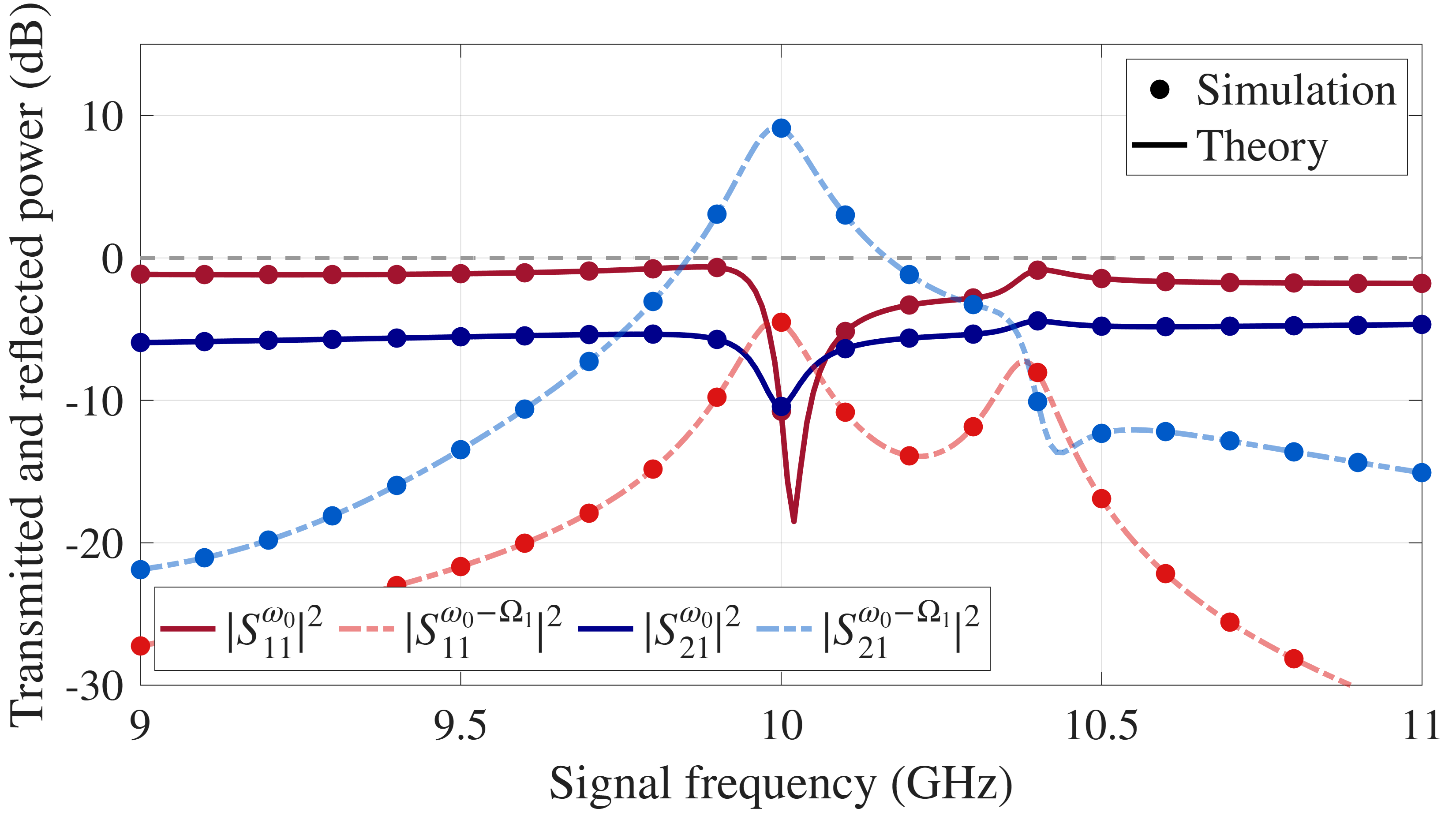}
\caption{{Transmitted and reflected power responses for a two-tone-modulated SDTWM loop network operating as a parametric amplifier. The dark and light blue lines correspond to the transmitted power at the signal frequency ($\omega_0$) for $m=0$ and the first idler frequency ($\omega_0-\Omega_1$) for $m=2$, respectively. The dark and light red lines correspond to reflected power at the signal and first idler frequencies, respectively. Dots correspond to results from harmonic balance simulations.}}
\label{fig:5}
\end{figure}

\begin{table}
\centering
\caption{Design parameters of the two-tone modulated SDTWM parametric amplifier}
\label{tab:2}
\vspace{0.5em}

\begin{tabularx}{0.4\textwidth} {|Y|Y|Y|Y|}
\hline
Tone ($\nu$) & $\Omega_\nu/2\pi$ & $\Omega_\nu t_\nu$ & $M_\nu$ \\
\hline
1 & 2.42\,GHz & $180^\circ$ & 0.123 \\
\hline
2 & 15.62\,GHz & $90^\circ$ & 0.227 \\
\hline
\end{tabularx}
\end{table}

The network's response is shown in Fig.~\ref{fig:5}.
These results demonstrate that a combination of frequency conversion and parametric amplification leads to gain at the first idler frequency ($\omega_0 - \Omega_1$).
Importantly, the network is now matched at the signal frequency ($\omega_0$), which resolves the strong reflection observed in the single-tone example of Section~\ref{sec:3_1_1}.
This highlights how multi-tone modulation can overcome limitations by enabling simultaneous coupling between multiple modes.
However, Fig.~\ref{fig:5} also shows that high reflection remains at the first idler frequency ($\omega_0 - \Omega_1$).
While the input is matched at the signal frequency, the reflected power at the idler remains below 0\,dB.
This indicates that the input resistance in the two-tone case is not negative, which differs from simple single-tone parametric amplification.

Both the simulation and the semi-analytical framework use $q = 11$ harmonics for each modulation tone. 
The time-varying capacitances in Keysight ADS were implemented based on the supplementary material of \cite{Sounas2013}, with the modification that, in this example, two modulation tones are used.

{To summarize, the two-tone modulation scheme enables simultaneous frequency conversion and parametric amplification, resulting in amplification at the down-converted frequency.
This also achieves input matching at the signal frequency, which resolves the limitation observed in the single-tone example of Section~\ref{sec:3_1_1}.
Reflection now appears at the idler frequency.
However, it does not exceed 0\,dB as in the single-tone case.
In the following section, we propose additional techniques to overcome the high reflection observed in this example.}
By adding conductors to the network, additional Bloch-Floquet modes are introduced.
As a result, the idler can be selected from a different conductor
With proper engineering, that idler mode can be isolated from the signal conductor, ensuring that no reflection occurs in the signal mode on the excited conductor.

\subsection{Multi-Conductor SDTWM Network} \label{sec:3_3}
{In this section, we extend the SDTWM loop formulation to multi-conductor networks that support multiple {LTI Bloch} modes. 
The key point is that the same driven modal framework developed for the two-conductor case still applies. 
The only change is that the eigenvectors and ABCD matrices must now index both the signal conductors and the frequency harmonics. 
Consequently, the formulation involves larger block matrices whose rows and columns correspond to the conductors, while each matrix entry is itself a sub-block that accounts for harmonic coupling.
As a result, the block-matrix formulation captures conductor-to-conductor modal coupling together with inter-harmonic coupling in a unified manner, while still allowing the response of the full loop network to be obtained from a single unit cell.}

In the time-invariant case, multi-conductor networks support multiple {LTI Bloch} modes.
The number of supported forward or reverse propagating modes is equal to the number of signal conductors.
As a result, increasing the number of signal conductors directly increases the number of supported {LTI Bloch} modes.
In practice, coupled transmission lines, overmoded waveguides, or leaky-wave antennas, supporting multiple {LTI} modes can be modeled with multi-conductor networks \cite{7756320}.
Generalizing such systems with space-time variation to realize multi-harmonic SDTWM networks requires a generalized framework for analysis.
Multi-conductor SDTWM networks provide a powerful platform for controlling both multiple {LTI} modes and frequency harmonics simultaneously.

\begin{figure}[!t]
\centering%
\subfloat[]{%
\centering
\includegraphics[width=50mm]{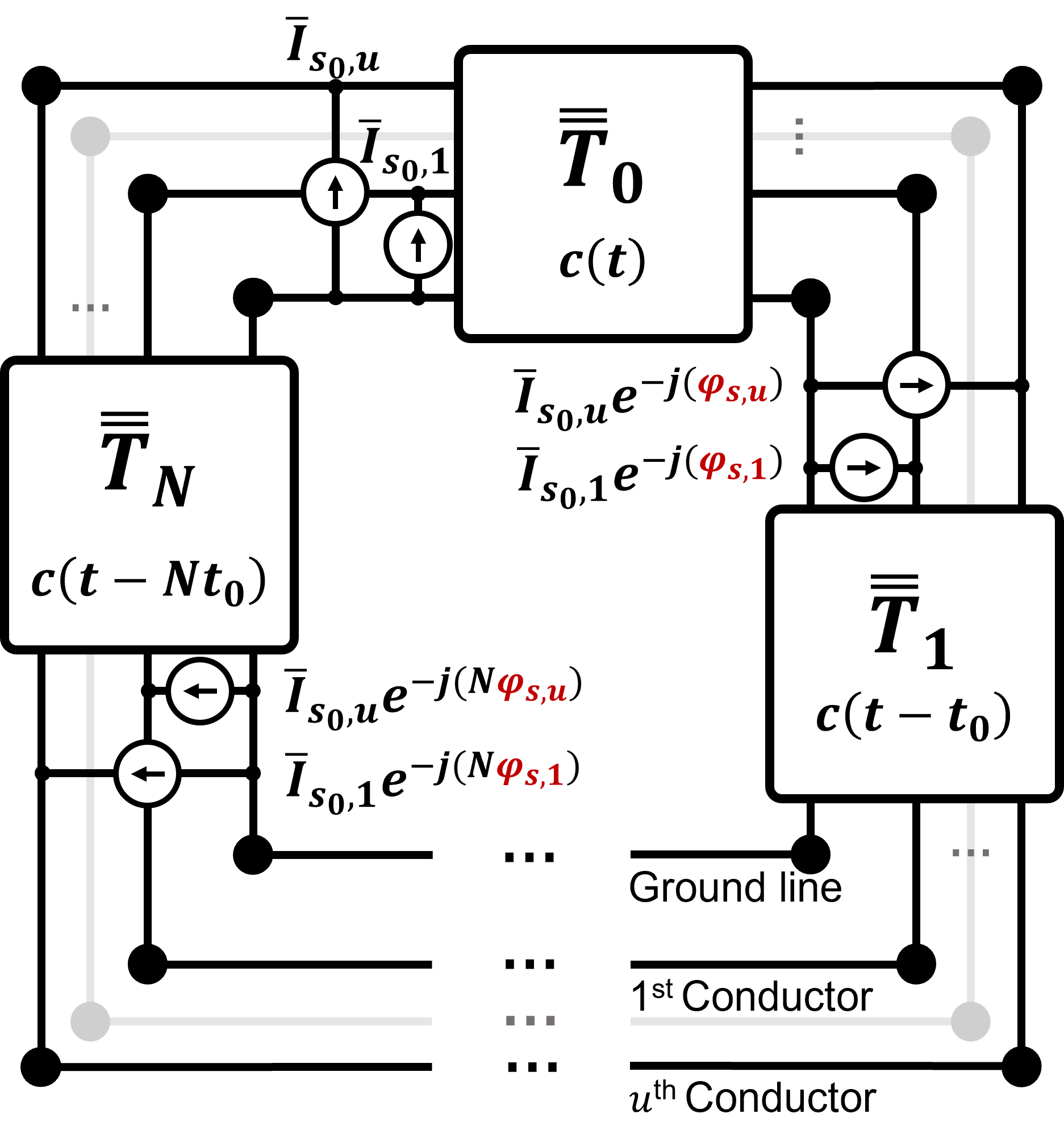}
\label{fig:6_a}
}%
\\[-0.1mm]%
\subfloat[]{%
\centering
\includegraphics[width=70mm]{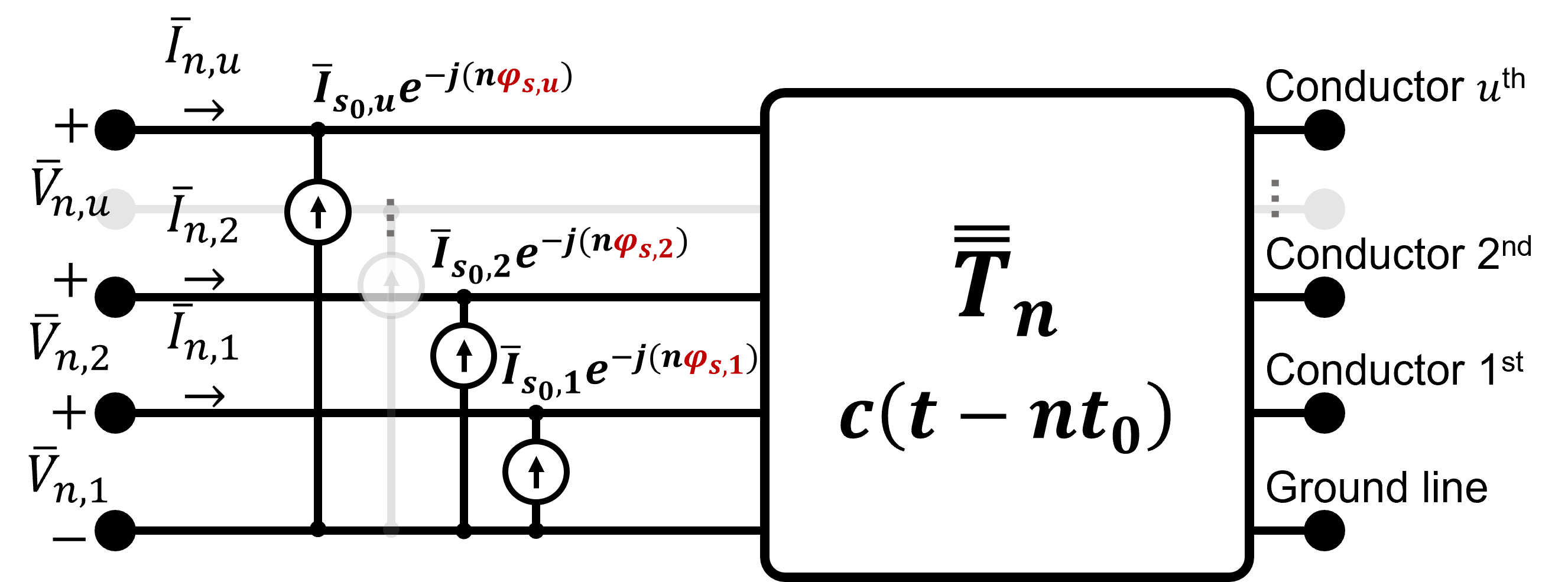}
\label{fig:6_b}
}%
\caption{Circuit schematic of a multi-conductor loop.  
(a) A $u$-conductor SDTWM loop network excited by phased array of $u$ sources, one for each conductor within each of the $N$ unit cells.  
(b) Representation of the $n^{\text{th}}$ unit cell, consisting of $(u)$ signal conductors and one ground line.}
\label{fig:6}
\end{figure}

Consider the $N$ unit cell multi-conductor SDTWM loop network shown in Fig.~\ref{fig:6_a}.  
The network contains $u+1$ conductors, where the zeroth conductor is the ground line, and conductors $1$ through $u$ serve as signal conductors.
Fig.~\ref{fig:6_b} shows the $n^{\text{th}}$ unit cell of this network.  
Each signal conductor can be excited by an independent phased array of sources, with their own phase delay and magnitude.  
This allows the excitation of different azimuthal orders.  
For instance, in a three-conductor system, the phased array on each signal conductor can be tailored to excite a different azimuthal order, allowing for modal diversity and enhanced control over wave propagation.

To find the driven modal solution of the $(u+1)$-conductor SDTWM loop network, let us start with the definition of the current sources at the $n^{\text{th}}$ unit cell, which is an extension of \eqref{eq:3} to multi-conductor systems:
\begin{equation}
\begin{bmatrix} 
\bar{I}_{s_n,1} \\
\bar{I}_{s_n,2} \\
\vdots \\
\bar{I}_{s_n,u} 
\end{bmatrix} =
\bar{\bar{\phi}}_s ^n
\begin{bmatrix} 
\bar{I}_{s_0,1} \\
\bar{I}_{s_0,2} \\
\vdots \\
\bar{I}_{s_0,u} 
\end{bmatrix}
\label{eq:8}
\end{equation}
Here, $\bar{I}_{s_0,u}$ is a vector representing the multi-harmonic source current on the $u^{\text{th}}$ conductor at the frequency harmonics.
The matrix $\bar{\bar{\phi}}_s$ accounts for the excitation phase delay applied to each conductor and is given by:
\begin{equation}
\bar{\bar{\phi}}_s =
\begin{bmatrix}
e^{j\varphi_{s,1}} \bar{\bar{I}} & \bar{\bar{0}} & \cdots & \bar{\bar{0}} \\
\bar{\bar{0}} & e^{j\varphi_{s,2}} \bar{\bar{I}} & \cdots & \bar{\bar{0}} \\
\vdots & \vdots & \ddots & \vdots \\
\bar{\bar{0}} & \bar{\bar{0}} & \cdots & e^{j\varphi_{s,u}} \bar{\bar{I}} \\
\end{bmatrix}
\label{eq:9}
\end{equation}
Here, $\varphi_{s,u}$ is the source phase delay applied to the $u^{\text{th}}$ conductor.
Each signal conductor has a single phase delay.
Note that the phase delay of the excitation on each conductor must satisfy the loop boundary condition given by \eqref{eq:4}.

By applying Kirchhoff’s Voltage Law (KVL) and Kirchhoff’s Current Law (KCL) at the input of the $n^{\text{th}}$ unit cell, we obtain:
\begin{equation}
\begin{bmatrix}
\bar{V}_{n,1} \\
\vdots \\
\bar{V}_{n,u} \\
\bar{I}_{n,1} \\
\vdots \\
\bar{I}_{n,u}
\end{bmatrix}
+
\begin{bmatrix} 
\bar{0} \\
\vdots \\
\bar{0} \\
\bar{I}_{s_n,1} \\
\vdots \\
\bar{I}_{s_n,u} 
\end{bmatrix}
=
\bar{\bar{T}}_n
\begin{bmatrix}
\bar{V}_{n+1,1} \\
\vdots \\
\bar{V}_{n+1,u} \\
\bar{I}_{n+1,1} \\
\vdots \\
\bar{I}_{n+1,u}
\end{bmatrix}
\label{eq:10}
\end{equation}
Here, $\bar{V}_{n,u}$ denotes the multi-harmonic total voltage at the input of the $n^{\text{th}}$ unit cell on the $u^{\text{th}}$ signal conductor.

\begin{figure}[!t]
\centering
\subfloat[]{%
\includegraphics[height=1.5in]{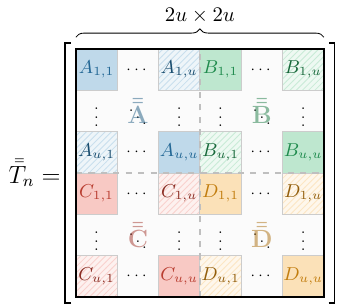}
\label{fig:Mat_3a}
}%
\vfill
\subfloat[]{%
\includegraphics[height=1.5in]{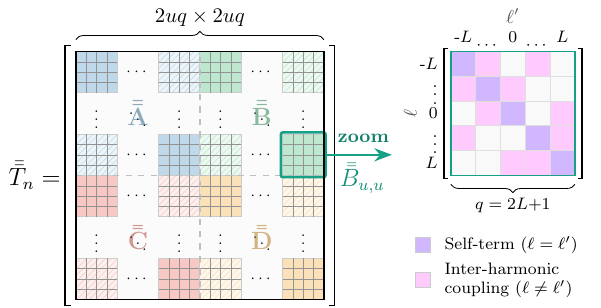}
\label{fig:Mat_3b}
}%
\caption{{ABCD matrices for (a) an LTI multi-conductor network and (b) a single-tone multi-conductor SDTWM network.}}
\label{fig:Mat_3}
\end{figure}

The ABCD matrix $\bar{\bar{T}}_n$ associated with the $n^{\text{th}}$ unit cell is defined as a $2 \times 2$ block matrix, given by:
\begin{equation}
\bar{\bar{T}}_n =
\begin{bmatrix}
\mathbf{\bar{\bar{A}}} & \mathbf{\bar{\bar{B}}} \\
\mathbf{\bar{\bar{C}}} & \mathbf{\bar{\bar{D}}}
\end{bmatrix}
\label{eq:11}
\end{equation}
{In the LTI case, each sub-block, such as $\bar{\bar{\mathbf{B}}}$, is a matrix of size $u \times u$ that accounts for modal coupling across $u$ signal conductors. 
In the special two-conductor example shown in Fig.~\ref{fig:Mat_1a}, this reduces to a $2 \times 2$ matrix.
This is shown in Fig.~\ref{fig:Mat_3a}. 
Each element $B_{i,j}$ is a scalar.
In the SDTWM network, shown in Fig.~\ref{fig:Mat_3b}, each element $\bar{\bar{B}}_{i,j}$ becomes a matrix that accounts for harmonic coupling across $q$ harmonics.
As a result, $\bar{\bar{\mathbf{B}}}$ has the following structure:}
\begin{equation}
\mathbf{\bar{\bar{B}}} =
\begin{bmatrix}
\bar{\bar{B}}_{1,1} & \cdots & \bar{\bar{B}}_{1,u} \\
\vdots & \ddots & \vdots \\
\bar{\bar{B}}_{u,1} & \cdots & \bar{\bar{B}}_{u,u}
\end{bmatrix}
\label{eq:A}
\end{equation}
Here, each element $\bar{\bar{B}}_{i,j}$ is itself a $q \times q$ block matrix, which accounts for the interaction between the $\ell^{\,\text{th}}$ and $\ell'^{\,\text{th}}$ frequency harmonics of the $i^{\text{th}}$ and $j^{\text{th}}$ conductors, respectively.
As a result, the overall ABCD matrix is of size $\bar{\bar{T}}_n \in \mathbb{C}^{2uq \times 2uq}$, reflecting the combination of conductors (modes) and frequency harmonics.
This framework, therefore, indexes over modes (conductor) and frequency harmonics.

Now, the Interpath Relation given by \eqref{eq:1} has been extended to a $(u+1)$-conductor network reads:
\begin{equation}
\begin{bmatrix}
\bar{V}_{n+1,1} \\
\vdots \\
\bar{V}_{n+1,u} \\
\bar{I}_{n+1,1} \\
\vdots \\
\bar{I}_{n+1,u}
\end{bmatrix}
=
\underbrace{\begin{bmatrix}
\bar{\bar{\phi_{s}}} & \bar{\bar{0}} \\
\bar{\bar{0}} & \bar{\bar{\phi_{s}}}
\end{bmatrix}}_{\bar{\bar{\Phi}}}
\bar{\bar{\mathcal{D}}}
\begin{bmatrix}
\bar{V}_{n,1} \\
\vdots \\
\bar{V}_{n,u} \\
\bar{I}_{n,1} \\
\vdots \\
\bar{I}_{n,u}
\end{bmatrix}
\label{eq:12}
\end{equation}
We define the matrix $\bar{\bar{\Phi}}$ to account for excitation phase delays, which are no longer scalar as in \eqref{eq:5}.
Additionally, both $\bar{\bar{\Phi}}$ and $\bar{\bar{\mathcal{D}}}$ matrices belong to the space $\mathbb{C}^{2uq \times 2uq}$.

By substituting \eqref{eq:12} into \eqref{eq:10} and performing straightforward algebraic manipulations, the driven modal solution for multi-conductor SDTWM networks is obtained as:
\begin{equation}
\begin{bmatrix}
\bar{V}_{n,1} \\
\vdots \\
\bar{V}_{n,u} \\
\bar{I}_{n,1} \\
\vdots \\
\bar{I}_{n,u}
\end{bmatrix}
=
\left(
\bar{\bar{T}}_n
\bar{\bar{\Phi}}
\bar{\bar{\mathcal{D}}}
- \bar{\bar{I}}
\right)^{-1}
\begin{bmatrix} 
\bar{0} \\
\vdots \\
\bar{0} \\
\bar{I}_{s_n,1} \\
\vdots \\
\bar{I}_{s_n,u} 
\end{bmatrix}
\label{eq:13}
\end{equation}
This expression provides the total voltage and current on each conductor in the network, accounting for both multi-modal and multi-harmonic interactions introduced by space-time modulation.

\subsubsection{Example 3: Electrically Small Antenna} \label{sec:3_3_1}

\begin{figure}[!t]
\centering%
\subfloat[]{%
\centering
\includegraphics[width=52mm]{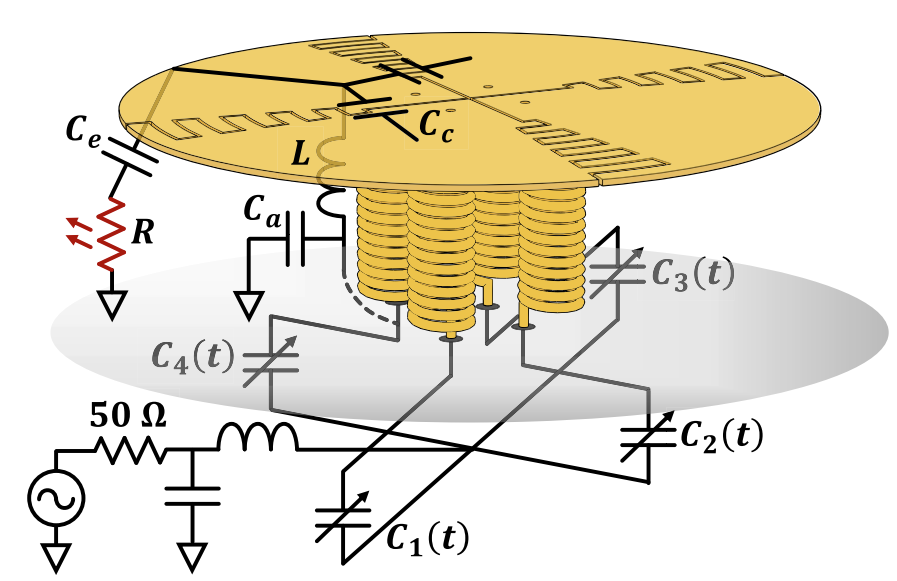}
\label{fig:7_a}
}%
\\[-0.1mm]%
\subfloat[]{%
\centering
\includegraphics[width=68mm]{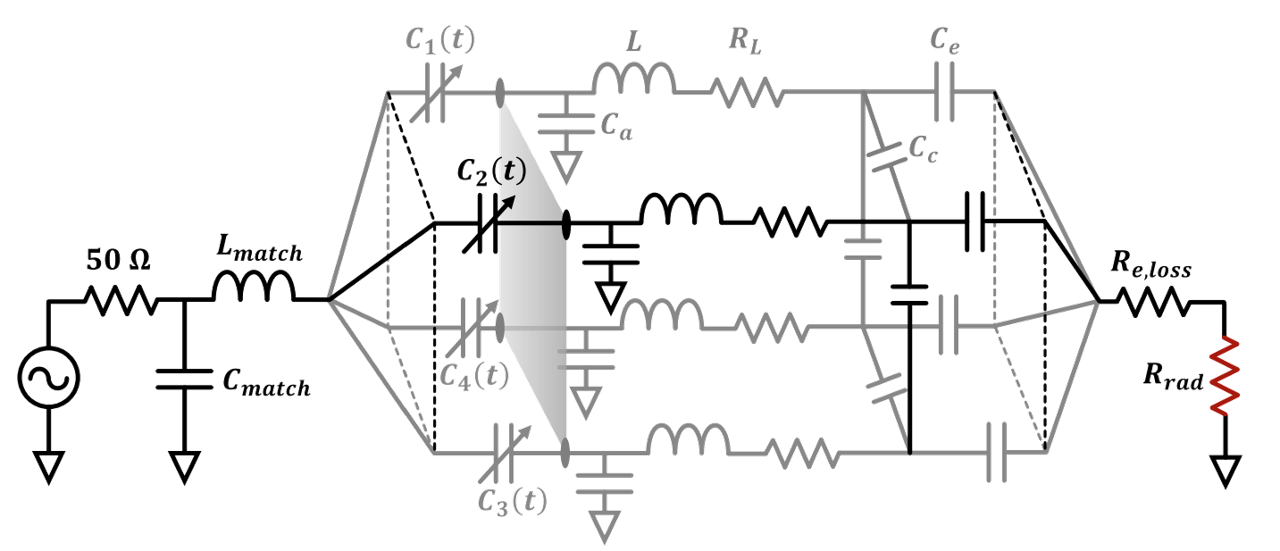}
\label{fig:7_b}
}%
\\[-0.1mm]%
\subfloat[]{%
\centering \hspace{2mm}
\includegraphics[width=68mm]{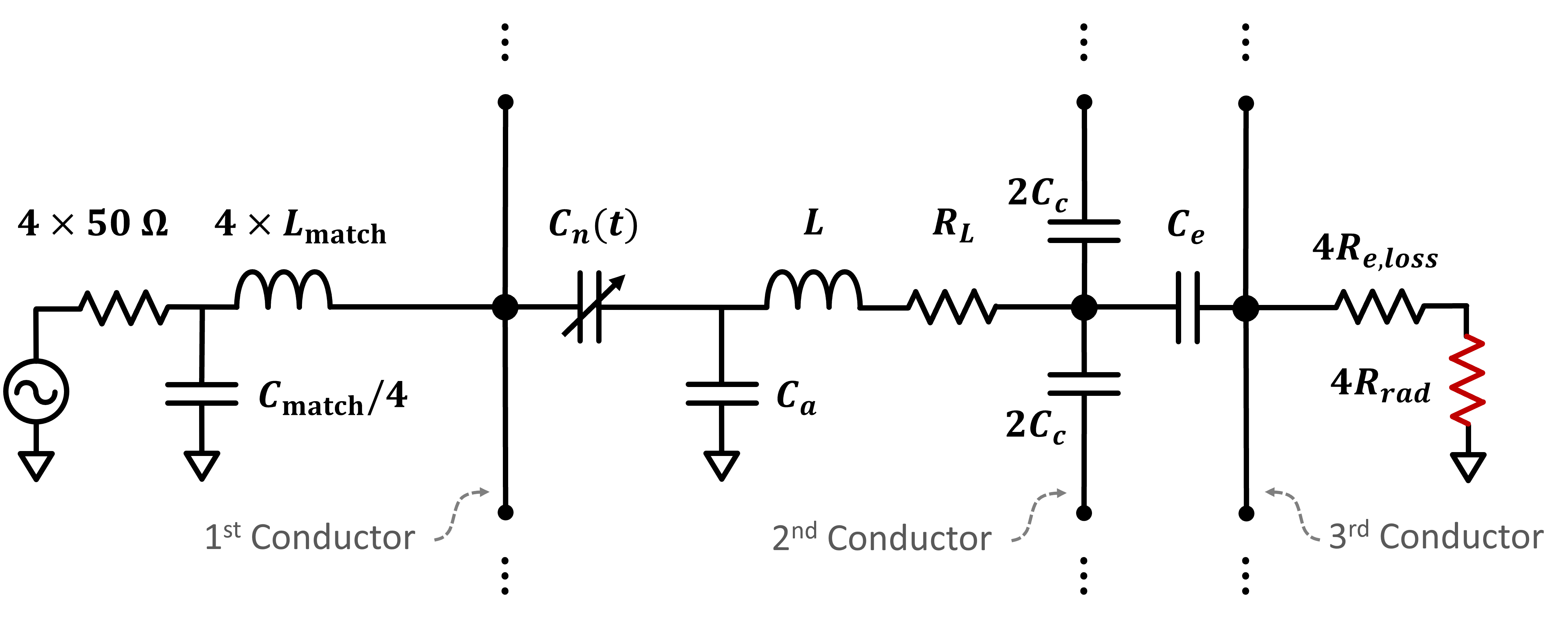}
\label{fig:7_c}
}%
\caption{Four-sector, top-hat-loaded monopole and its equivalent circuit model.  
(a) Antenna geometry with circuit elements overlaid: interdigitated capacitors ($C_c$) couple neighboring sectors, helical inductors ($L$) reduce electrical size ($ka = 0.253$ at 341\,MHz), and time-varying capacitors $C_1(t)$–$C_4(t)$ time modulate each sector.  
(b) Full four-path equivalent circuit with L-matching network. Only the in-phase ($m=0$) current mode excites the radiation resistance $R_{\text{rad}}$.  
(c) Unit-cell representation as a four-conductor SDTWM network including $L_{\text{match}}$, $C_{\text{match}}$, inter-sector coupling, and loss elements.}
\end{figure}

An illustrative example of a multi-conductor SDTWM network is the circuit equivalent of the electrically small antenna that the authors presented in~\cite{10785556}.
The antenna, shown in Fig.~\ref{fig:7_a}, consists of a top-hat loaded monopole segmented into four sectors.
In each sector, the vertical monopolar element is twisted into a helix for inductive loading.
This reduces its electrical size to $ka = 0.253$ at the fundamental self-resonant frequency of 341\,MHz, placing it within the electrically small regime \cite{6192311, Harrington1960}.
Each sector is modulated using time-varying capacitors distributed in series with the vertical element, forming a four-sector, SDTWM loop network with two-tone modulation.
The modulation waveform of the antenna follows the form given by \eqref{eq:7}.
The four feeds of the antenna are tied together, with the feed positioned at the center.

\begin{figure}[!b]
\centering
\includegraphics[width=95mm]{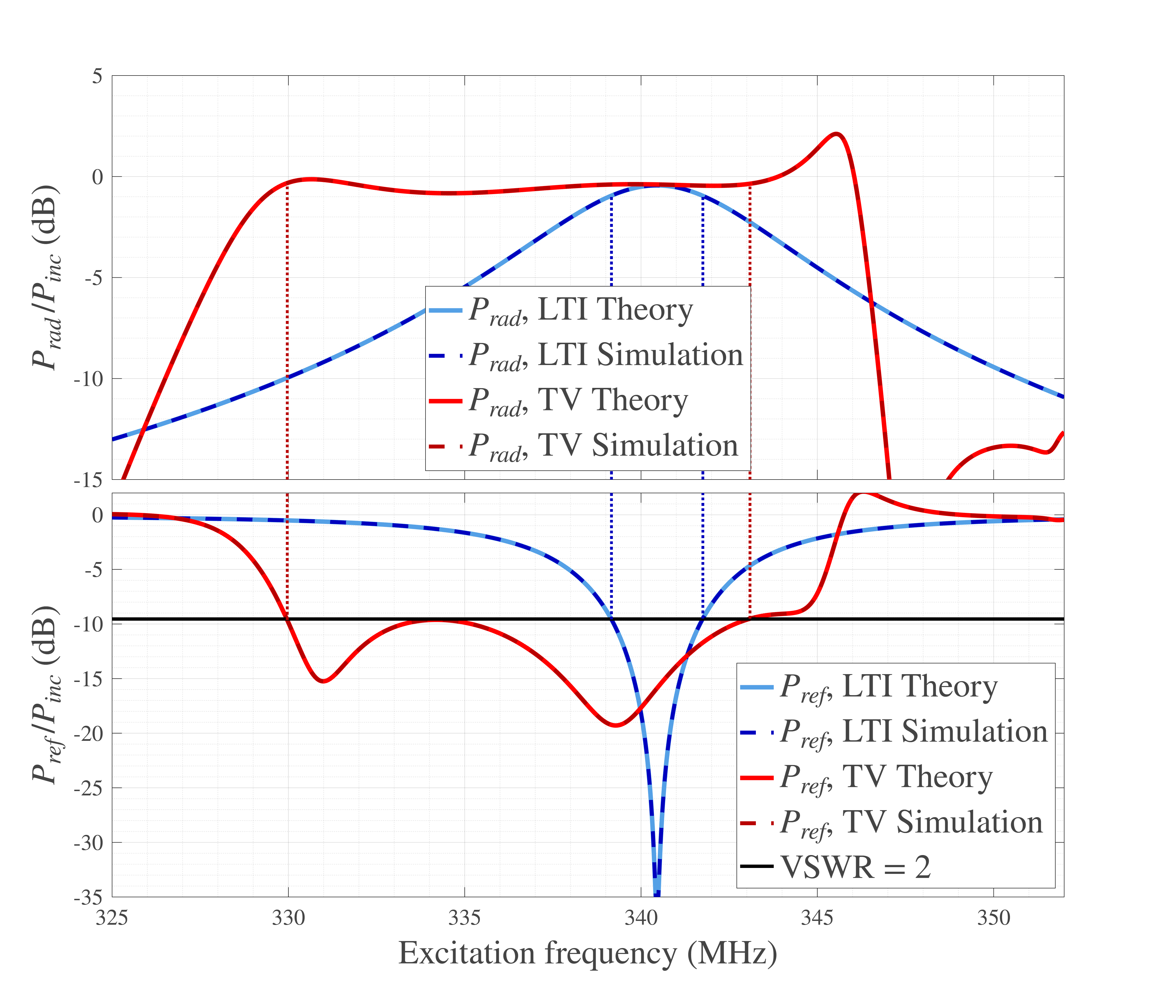}
\caption{Comparison of the LTI and two-tone time-varying antenna performance based on the theoretical circuit model and harmonic balance simulation. Top: Radiated power ($P_{\text{rad}}/P_{\text{inc}}$) versus frequency for the LTI (blue) and time-varying (red) antennas. Bottom: Reflected power ($P_{\text{ref}}/P_{\text{inc}}$), including VSWR = 2 threshold. Solid lines denote proposed circuit-based theory, while dashed lines represent harmonic balance simulations of the equivalent model. Time-varying modulation improves both the radiation bandwidth and matching performance.}
\label{fig:8}
\end{figure}

Using the modulation scheme provided in Table~\ref{tab:3}, where by exciting the zeroth azimuthal order ($\varphi_s = 0^\circ$) on the first conductor, the $m = 0$ mode is coupled to both $m = 2$ and $m = 3$ at negative frequencies, a matching bandwidth improvement greater than 6\,dB (under {VSWR = 2}) is achieved compared to the LTI version of the antenna, as shown in Fig.~\ref{fig:8}.
These results correspond to the antenna design presented in~\cite{10785556}.
This study further demonstrated that by detuning the modulation frequencies of both tones, a trade-off between gain and impedance matching bandwidth can be engineered.
In this case, the gain from the parametric process is sacrificed to achieve a broader matching bandwidth.

An equivalent circuit model is shown in Fig.~\ref{fig:7_b} \cite{10785556,6143985}.
It consists of an air-core inductor $L$, a parasitic capacitance between the feed and the ground plane $C_a$, an inter-sector coupling capacitance $C_c$, a fringing capacitance between the top hat and the ground plane $C_e$, and a radiation resistance $R_\text{rad}$.
{The ohmic loss of the air-core inductor, represented by $R_L$, is also included in the model. 
Since the proposed semi-analytical framework applies to networks composed of LPTV elements, practical effects such as loss can be incorporated in the unit-cell circuit model.}
The circuit parameters for the equivalent circuit model of the electrically small antenna are given in Table~\ref{tab:3}.
The unit cell of this equivalent circuit model is shown in Fig.~\ref{fig:7_c}, where a 4-conductor configuration is shown, comprising three signal conductors and one ground conductor.
Here, the first signal conductor represents the tied feeds on the source side, the second signal conductor models the top-hat antenna, and the third signal conductor represents the radiation resistance.

In the $m = 0$ mode, the sectors (unit cells) are in phase ($\varphi_s = 0$) and the antenna radiates efficiently.
In contrast, the other three modes, which have different resonant frequencies, exhibit destructive interference due to the vectorial near cancellation of their sector currents.
This behavior mimics the functionality of an N-path filter \cite{8233412, PhysRevApplied.14.064060}.
Hence, the multi-conductor network model effectively captures the N-path topology.

Now, by taking a single unit cell (sector) of the antenna, shown in Fig.~\ref{fig:7_c}, and combining the analysis presented in Sections~\ref{sec:3_2} and~\ref{sec:3_3} for multi-tone and multi-conductor SDTWM networks, the driven modal solution for this antenna can be obtained. 
The first step is to formulate the ABCD matrix of a unit cell.
For a network with $v$ modulation tones and $u$ conductors, the ABCD matrix is given by $\bar{\bar{T}}_n \in \mathbb{C}^{2uq^v \times 2uq^v}$.
In this particular case, with 4 conductors ($u=3$) and $v = 2$ tones, we have $\bar{\bar{T}}_n \in \mathbb{C}^{6q^2 \times 6q^2}$, assuming $q$ harmonics are included in the analysis.

\begin{table}
\centering
\caption{Equivalent circuit parameters for the multi-conductor SDTWM electrically small antenna}
\label{tab:3}
\vspace{0.5em}

\begin{tabularx}{0.44\textwidth}{|Y|Y|Y|Y|Y|Y|}
\hline
$L_\text{match}$ & $C_\text{match}$ & $L$ & $C_a$ & $C_e$ & $C_c$ \\
\hline
4.42\,nH & 47.2\,pF & 291\,nH & 0.45\,pF & 0.81\,pF & 1.35\,pF \\
\hline
\end{tabularx}

\vspace{0.2em}

\begin{tabularx}{0.44\textwidth}{|Y|Y|Y|Y|Y|}
\hline
$R_L$ & $R_\text{e,loss}$ & $R_\text{rad}$ & $C_0$ & $\varphi_s$ \\
\hline
0.69\,$\Omega$ & 0.186\,$\Omega$ & 1.434\,$\Omega$ & 10\,pF & $0^\circ$ \\
\hline
\end{tabularx}

\vspace{0.6em}

\begin{tabularx}{0.4\textwidth}{|Y|Y|Y|Y|}
\hline
Tone ($\nu$) & $\Omega_\nu/2\pi$ & $\Omega_\nu t_\nu$ & $M_\nu$ \\
\hline
1 & 486.24\,MHz & $180^\circ$ & 0.3 \\
\hline
2 & 538.88\,MHz & $90^\circ$ & 0.17 \\
\hline
\end{tabularx}
\end{table}

By using the driven modal solution given by~\eqref{eq:13}, the voltages on all signal conductors can be calculated.
The voltage on the third conductor can be used to compute the power through $R_{rad}$, which represents the power transmitted by the antenna.
Similarly, the voltage on the first conductor can be used to compute the reflected power at the input.
As illustrated in Fig.~\ref{fig:7_c}, the antenna is excited using a voltage source.
To remain consistent with the driven modal solution developed in this work, an equivalent Norton representation can be used, where a current source feeds the first signal conductor.
A comparison between the proposed analysis framework and harmonic balance simulations, shown in Fig.~\ref{fig:8}, demonstrates excellent agreement.

{To summarize, in this example, a multi-conductor SDTWM network is used to create an N-path filtering effect that suppresses unwanted frequency harmonics at both the input and radiated.
In addition, the two-tone modulation scheme is designed to operate as a parametric amplifier.
This configuration can be engineered to trade gain for improved impedance matching, resulting in wideband matching across the relevant harmonics and suppression of idler radiation at the output \cite{10785556}.
As a result, in contrast to the examples in Sections~\ref{sec:3_1_1} and~\ref{sec:3_2_1}, no reflections are observed at any idler frequencies.
This example highlights the utility of the proposed framework for modeling multi-conductor SDTWM networks.}

\subsection{Computational Complexity}
{The computational cost of the proposed framework is determined by solving the linear system in \eqref{eq:6}. 
For a $v$-tone modulation with $q$ harmonics per tone, the number of spectral components is $q^v$. 
For a network with $u$ conductors, the matrices in \eqref{eq:6} have dimensions $(2u q^v \times 2u q^v)$. 
Using a direct solver, the computational cost scales as $\mathcal{O}((2u q^v)^3)$ and the memory requirement as $\mathcal{O}((2u q^v)^2)$. 
The dominant factor is the number of modulation tones $v$, since the spectral dimension grows exponentially as $q^v$, while the number of conductors affects the matrix size only linearly.}

{For comparison, harmonic balance (HB) simulations solve a nonlinear system of $K q^v$ equations, where $K$ is the circuit size (the number of independent node voltages and branch currents). 
Using a direct Newton solver, the computational complexity scales approximately as $\mathcal{O}((K q^v)^3)$ with memory scaling as $\mathcal{O}((K q^v)^2)$. 
Since HB includes the entire circuit network, $K$ increases with the number of unit cells and the internal unit cell circuit complexity. 
In contrast, the proposed semi-analytical formulation analyzes a single unit cell and uses only the terminal voltages and currents as unknowns, so the system dimension does not grow with the number of cascaded unit cells.}

{As an example, the electrically small antenna shown in Fig.~\ref{fig:7_b} contains $K=16$ independent nodes. 
Thus, the HB complexity scales as $\mathcal{O}((16 q^v)^3)$. 
The proposed framework models the antenna as a three-conductor unit cell ($u=3$) shown in Fig.~\ref{fig:7_c}, giving a matrix dimension of $6 q^v$ and a complexity of $\mathcal{O}((6 q^v)^3)$. 
For the same number of harmonics per tone, this reduces the computational cost by approximately a factor of nineteen. 
For $q=11$, the MATLAB implementation requires 6.54\,s, whereas the ADS harmonic balance solver requires 120\,s, corresponding to an approximately $18.35\times$ speed improvement. 
All simulations were performed on a desktop computer with an Intel Core i7-12700 CPU (2.10\,GHz) and 64\,GB RAM.}

{The primary computational limitation of the proposed approach remains the exponential growth of the spectral dimension with the number of modulation tones $v$. 
This can be mitigated by limiting the harmonic truncation of each tone or by using faster matrix multiplication algorithms (e.g., Alman--Williams), which can reduce the theoretical complexity from $\mathcal{O}(n^3)$ to approximately $\mathcal{O}(n^{2.37})$.}

{In practice, the parameters $u$ and $v$ are set by the physical SDTWM network and the intended modulation functionality, whereas $q$ is a truncation parameter chosen for accuracy. 
Specifically, $u$ is determined by the number of conductors, or equivalently, the number of {LTI Bloch} modes that must be included in the circuit model. 
Similarly, $v$ is determined by the number of modulation tones required to realize the desired simultaneous space-time couplings. 
In contrast, $q$ determines the number of considered frequency harmonics in the truncated multi-harmonic formulation. 
Its required value depends mainly on the modulation waveform and the modulation depth $M$. 
For example, the cosine modulation used in this work mainly couples adjacent harmonics, so relatively few harmonics are often sufficient, whereas waveforms with richer Fourier content generally require a larger $q$. 
Likewise, for small modulation depths, only limited energy is transferred to higher-order harmonics, so a small truncation order is often adequate, while larger modulation depths require more harmonics to be considered. 
In practice, $q$ is selected by increasing the number of considered harmonics until the solution converges.}

\section{Spatial Green's Function of a Loop SDTWM Network} \label{sec:4}
Traditional impedance and scattering parameters are formulated for linear time-invariant (LTI) systems, where each frequency component of the signal is independent of other frequencies.
In SDTWM networks, however, the space-time modulation couples different harmonics, causing energy exchange between frequencies.
As a result, conventional LTI tools fail to capture inter-harmonic interactions and frequency conversion.
To address this, we employ a spatial Green’s function formulation that inherently describes the spatiotemporal response of the SDTWM network, relating inputs to outputs across all interacting harmonics.
\subsection{Analytic Array Scanning} \label{sec:4_1}

\begin{figure}[!t]
\centering
\includegraphics[width=70mm]{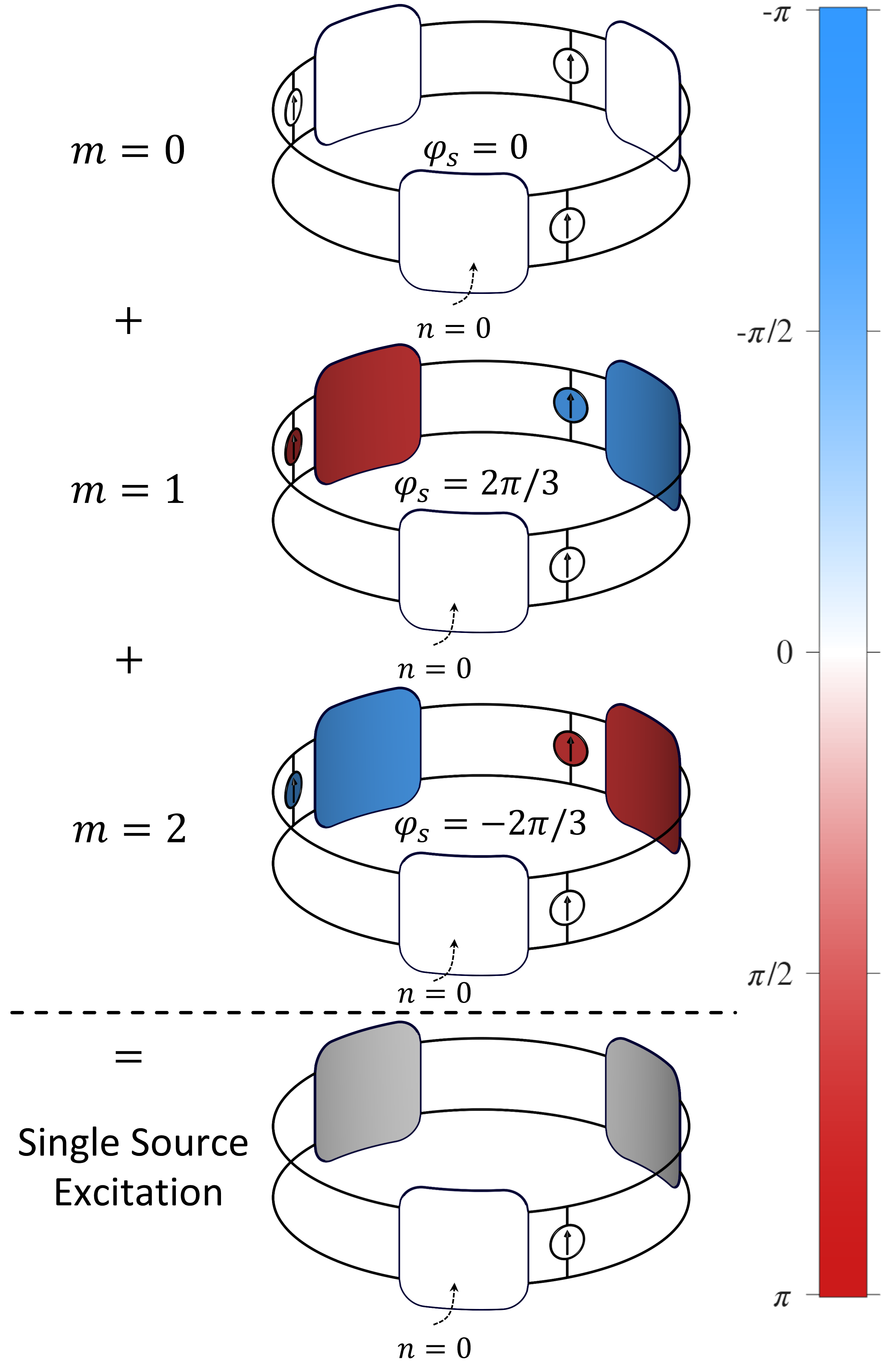}
\caption{Visualization of the spatial Green’s function in SDTWM loop networks using the Analytic Array Scanning (AAS) method.
Each loop network consists of $N=3$ unit cells and is excited by a phased array of current sources with azimuthal order $m$, where $\varphi_s = \frac{2\pi m}{N}$.
Source phases are color-coded, ranging from white (zero phase) to red (positive phase) and blue (negative phase).
The zeroth unit cell is shown in white, as it is assigned zero phase by the definition.
The single-source excitation (SSE) response is obtained by summing these modal solutions, where all current sources cancel out vectorially due to orthogonality, except the one at the zeroth unit cell.}
\label{fig:SSE}
\end{figure}

Based on the methodology introduced in~\cite{Babaee2024}, and inspired by work on LTI spatially periodic structures~\cite {doi:10.1080/02726349908908643, 1256761}, an Analytic Array Scanning (AAS) technique is proposed for computing the spatial Green’s function of SDTWM loop networks.
In SDTWM loop networks with $N$ unit cells, each driven modal solution corresponds to a phased excitation with delay $\varphi_s = \frac{2\pi m}{N}$, where $m = 0, 1, \dots, N{-}1$, as defined by~\eqref{eq:4}.
By summing the driven modal solutions derived using~\eqref{eq:6}, the total voltages and currents due to a single source excitation {(SSE)} are obtained as:
\begin{equation}
\begin{bmatrix} \bar{V}_{n} \\ \bar{I}_{n} \end{bmatrix}^{SSE} = \frac{1}{N} \sum_{m=0}^{N-1}
\left( e^{-j m \frac{2\pi}{N}} \bar{\bar{T}}_{n} \bar{\bar{\mathcal{D}}}-\bar{\bar{I}} \right)^{-1}
\begin{bmatrix} \bar{0} \\ \bar{I}_{s_n} \end{bmatrix}
\label{eq:14}
\end{equation}

The vectorial superposition cancels all current sources except for the zeroth source due to the orthogonality relation,
\begin{equation}
\frac{1}{N} \sum_{m=0}^{N-1} e^{j n \frac{2\pi m}{N}} =
\begin{cases}
1, & n = 0 \\
0, & n \neq 0~.
\end{cases}
\label{eq:15}
\end{equation}

The single-source excitation (SSE) solution in~\eqref{eq:14} is derived using the driven modal solutions for single or multi-tone modulation, given by~\eqref{eq:6}.
The AAS method can also be applied to multi-conductor SDTWM networks using the driven modal formulation given by~\eqref{eq:13}.

Fig.~\ref{fig:SSE} illustrates the single-source excitation (SSE) response described by \eqref{eq:14} for a three-unit-cell SDTWM loop network.
It illustrates various azimuthal orders $m \in \{ 0, 1, 2\}$, with progressive phase $\varphi_s = \frac{2\pi m}{N}$.
While all unit cells’ current sources are phase-shifted according to the azimuthal order, the zeroth unit cell remains at zero phase (white).
Upon summation of the modal solutions, only the zeroth unit cell retains a current source, while all others cancel out vectorially due to orthogonality.

\subsubsection{{Example 4: Non-magnetic Circulator}} \label{sec:4_2}
\begin{figure}[!b]
\centering%
\subfloat[]{%
\centering
\includegraphics[width=50mm]{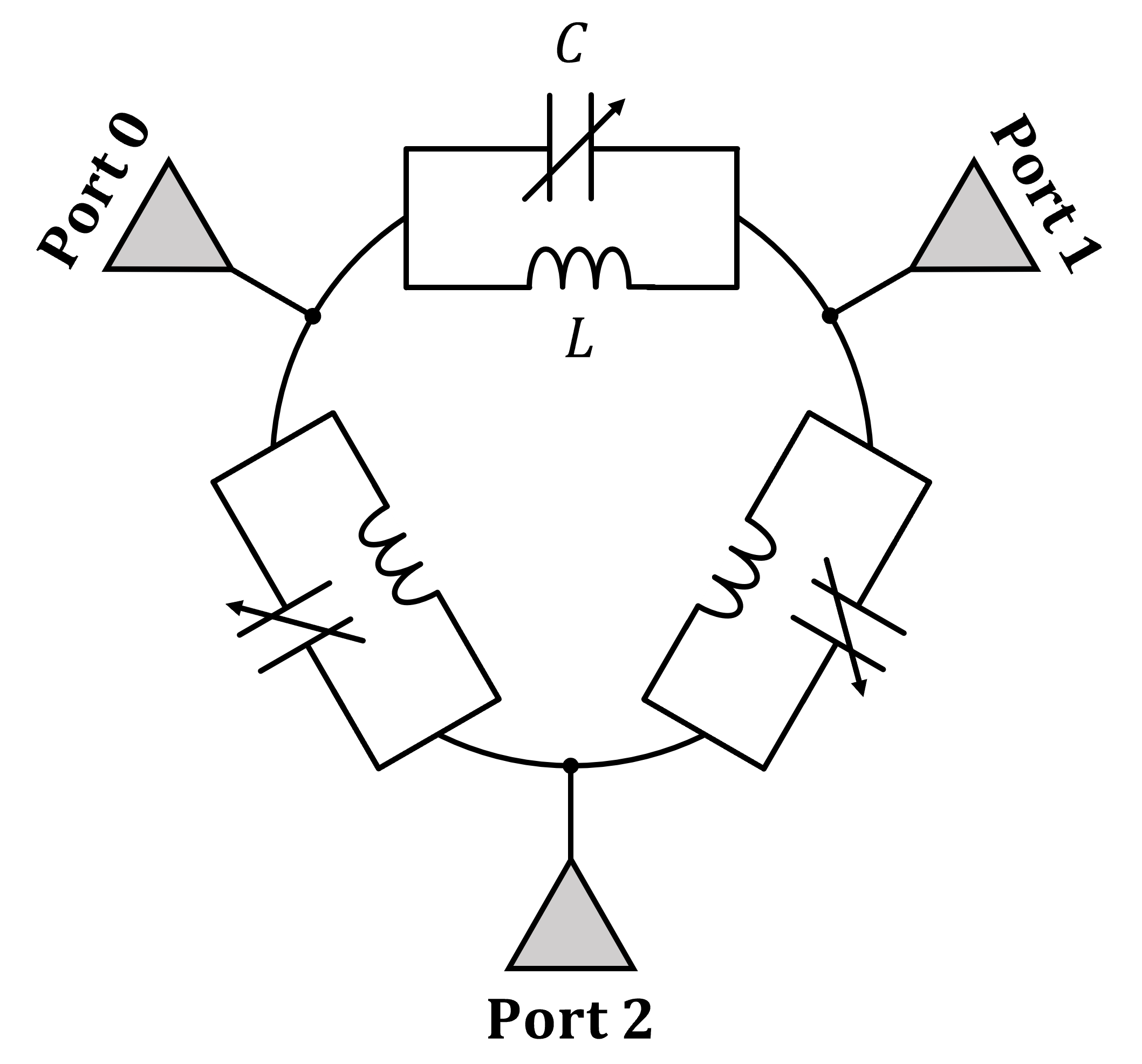}
\label{fig:9_a}
}%
\\[-0.1mm]%
\subfloat[]{%
\centering
\includegraphics[width=50mm]{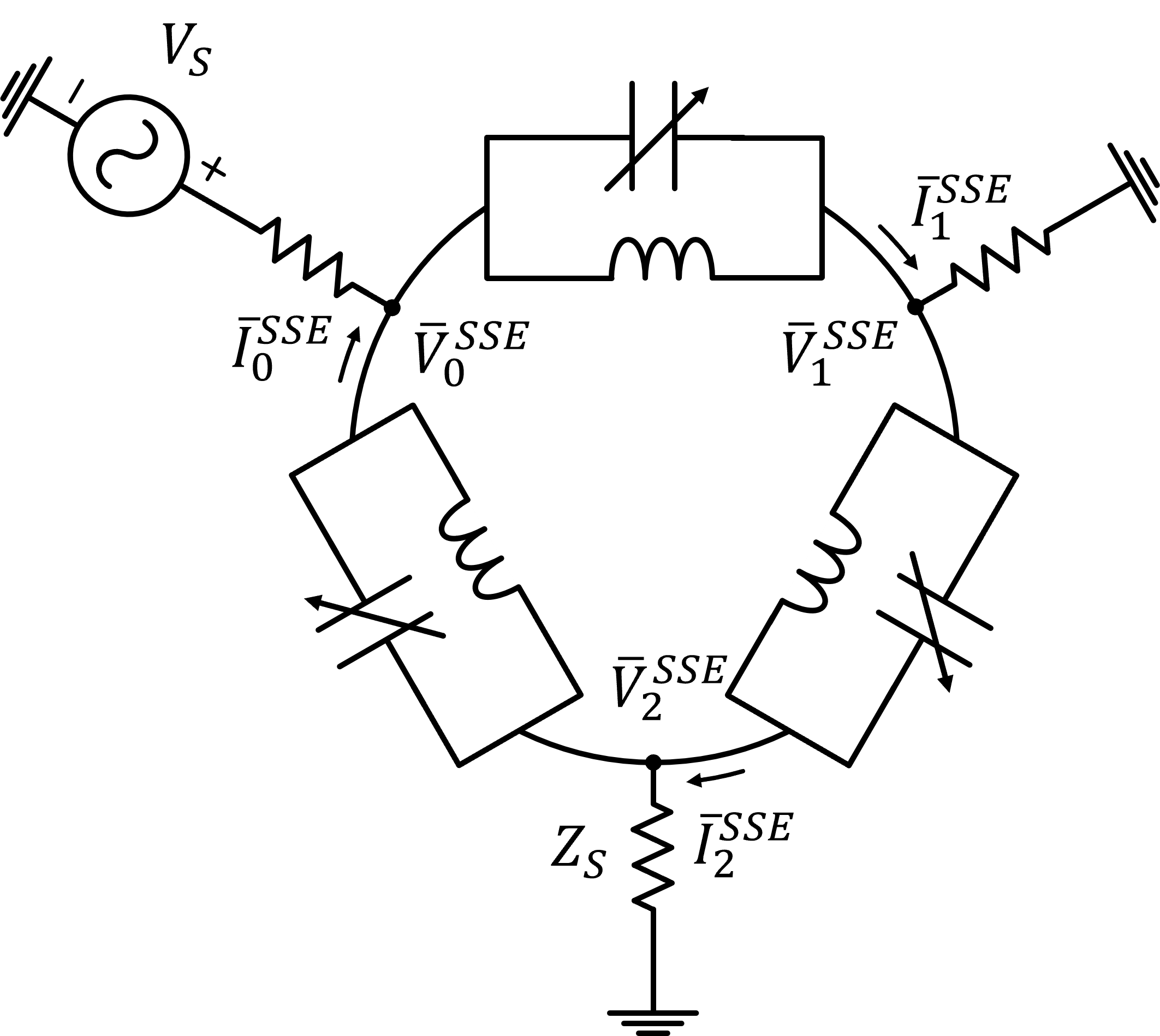}
\label{fig:9_b}
}%
\\[2.6mm]
\caption{(a) The non-magnetic space-time modulated circulator proposed in \cite{9257419} (b) Representation of the circulator with single-source excitation.}
\label{fig:9}
\end{figure}

The Green's function of SDTWM networks will be used to design the three-port circulator \cite{Babaee2025IMS} shown in Fig.~\ref{fig:9_a}.
The unit cell comprises a parallel LC resonator with a shunt source and source resistance.
Here, the total number of unit cells is $N=3$.
The capacitance of the $n^{\text{th}}$ unit cell is modulated using a traveling-wave single-tone modulation described by~\eqref{eq:7}.

The circulator can be analyzed under a single-source excitation (see Fig.~\ref{fig:9_b}), using (\ref{eq:14}).
The source is placed at port 0, and the aim is to match the source to the network at the operating frequency.
Additionally, all signal power should be directed to port 1, while port 2 remains isolated.
In other words, in Fig.~\ref{fig:9_b}, $\bar{V}_{0}$ should be half of the source voltage given that port 0 is impedance matched.
Voltage $\bar{V}_{1}$ should also be half of the source voltage since the input signal is directed to port 1.
Finally, $\bar{V}_{2}$ should be zero, given that port 2 is isolated.
These conditions can be expressed as follows,
\begin{equation}
\begin{bmatrix} \bar{V}_{0} \\ \bar{V}_{1} \\ \bar{V}_{2} \end{bmatrix}^{SSE} =
\begin{bmatrix} \frac{\bar{V}_{S}}{2} \\ \frac{\bar{V}_{S}}{2} \\ \bar{0} \end{bmatrix}.
\label{eq:16}
\end{equation}
Here, $\bar{V}_{S}$ is a vector representing the voltage source magnitudes across different frequency harmonics.
Voltage magnitudes are set to zero for all harmonics except at the fundamental frequency ($\omega_0/2\pi$), where its value is equal to one.

\begin{figure}
\centering%
\subfloat[]{%
\centering
\includegraphics[width=90mm]{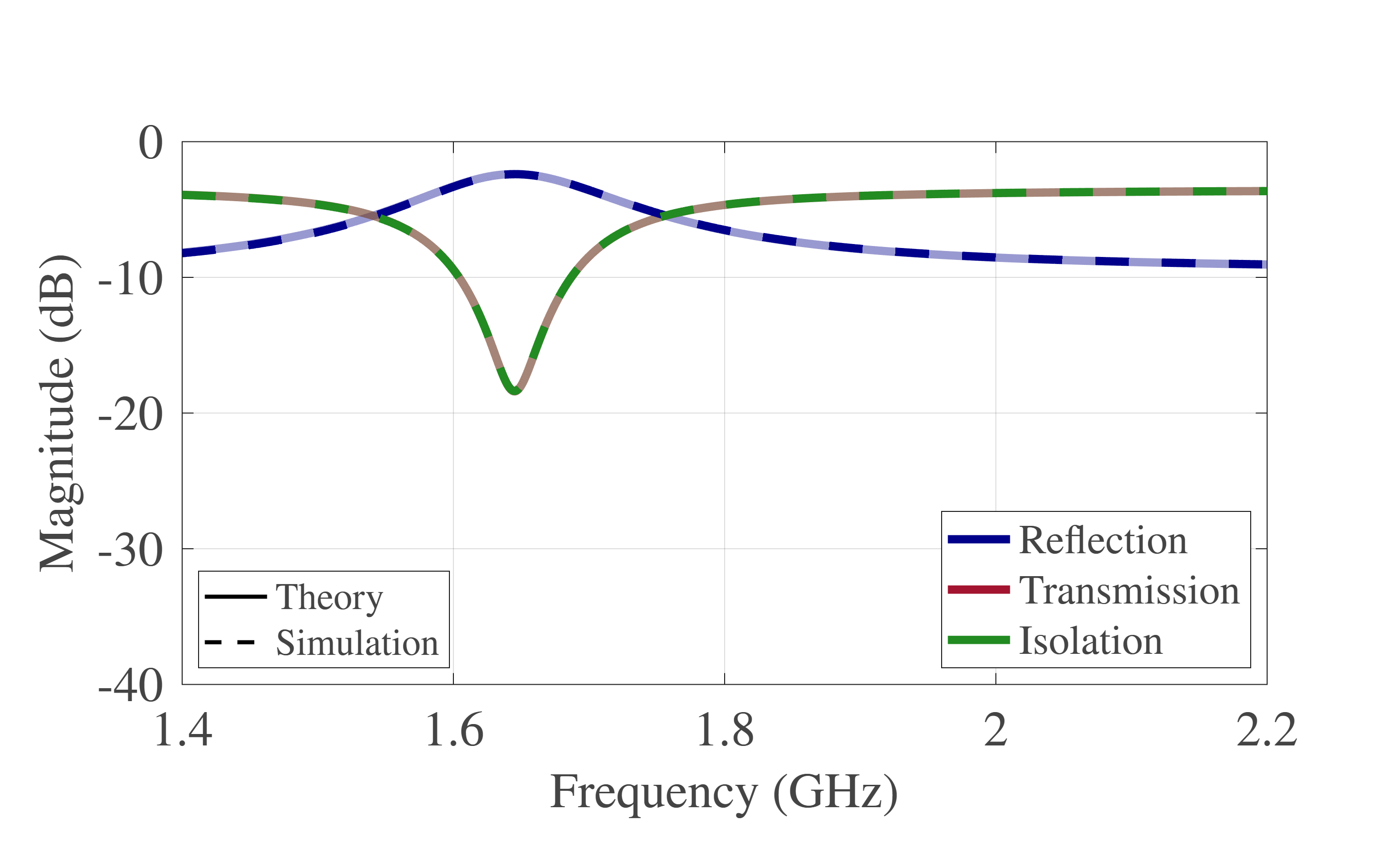}
\label{fig:10_a}
}%
\\[-0.1mm]%
\subfloat[]{%
\centering
\includegraphics[width=90mm]{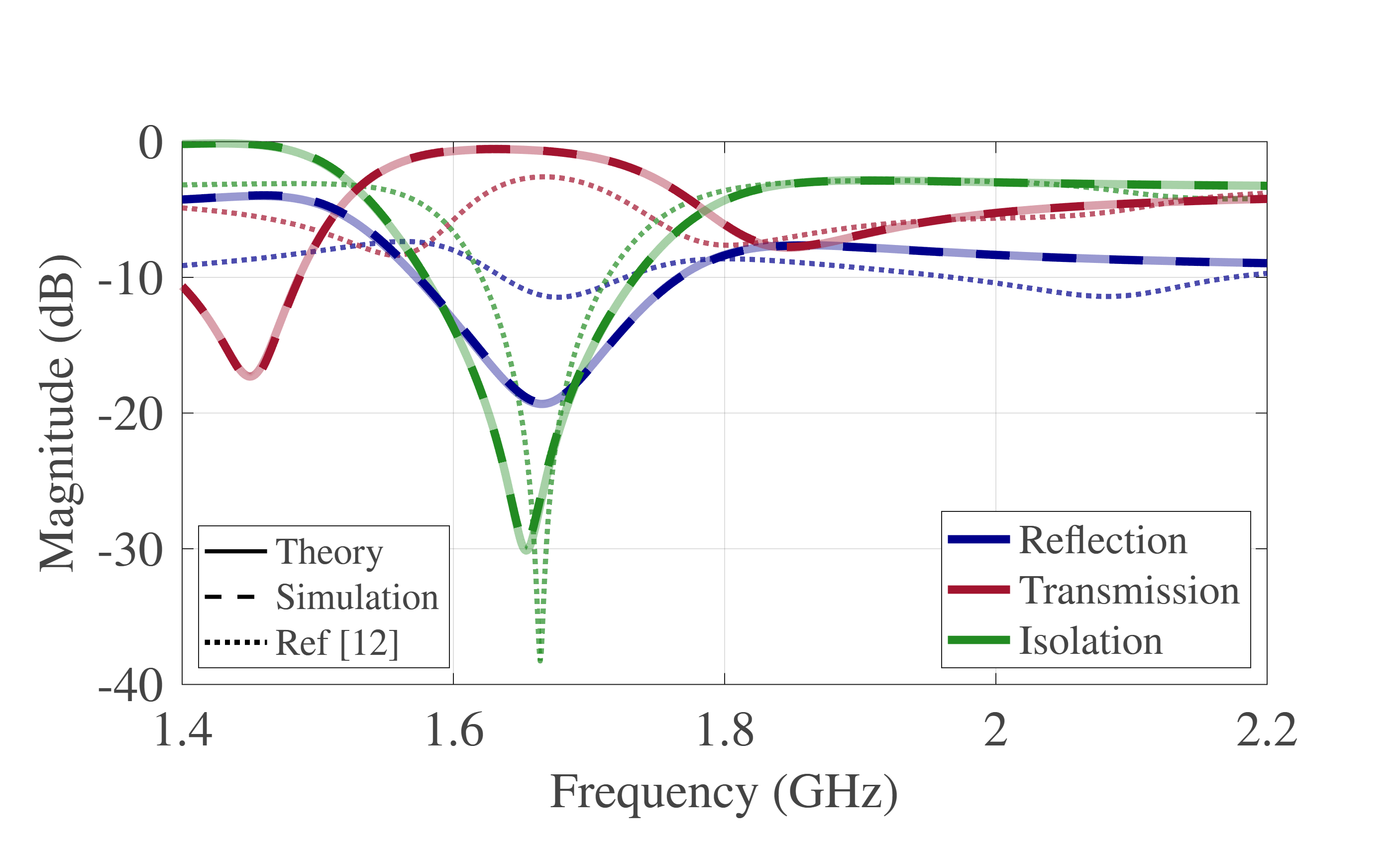}
\label{fig:10_b}
}%
\\[2.6mm]
\caption{S-parameters of the circulator
(a) without modulation and (b) with modulation.}
\label{fig:10}
\end{figure}

\begin{table}
\begin{center}
\centering
\caption{Design parameters for the SDTWM loop network designed to be a circulator.}
\label{tab:4}
\scalebox{1.1}{
\begin{tabular}{| c | c | c | c |}
\hline
$\omega_0/{2\pi}$ & $\Omega/{2\pi}$ & $\Omega t_0$ & $M$ \\
\hline
$1.66 \,$GHz & $2.77 \,$GHz & $120^\circ$ & $0.35$ \\
\hline\hline
$C$ & $L$ & $Q$ & $Z_s$ \\
\hline
$7.8 \,$pF & $1.2 \,$nH & $55$ & $50 \, \Omega$ \\
\hline
\end{tabular}}
\end{center}
\end{table}

As described in Section~\ref{sec:4_1}, a single-source excitation can be decomposed into three modal solutions, allowing (\ref{eq:16}) to be rewritten in terms of the driven modal voltages of the zeroth unit cell as follows,
\begin{equation}
\begin{bmatrix} \frac{\bar{V}_{S}}{2} \\ \frac{\bar{V}_{S}}{2} \\ \bar{0} \end{bmatrix} =
\frac{1}{3} 
\begin{bmatrix}
\bar{\bar{I}} & \bar{\bar{I}} & \bar{\bar{I}} \\
\bar{\bar{\mathfrak{D}}} & e^{-j\frac{2\pi}{3}}\bar{\bar{\mathfrak{D}}} & e^{j\frac{2\pi}{3}}\bar{\bar{\mathfrak{D}}} \\
\bar{\bar{\mathfrak{D}}}^2 & e^{j\frac{2\pi}{3}}\bar{\bar{\mathfrak{D}}}^2 & e^{-j\frac{2\pi}{3}}\bar{\bar{\mathfrak{D}}}^2 \\
\end{bmatrix}
\begin{bmatrix} \bar{V}_{0}^{m=0} \\ \bar{V}_{0}^{m=1} \\ \bar{V}_{0}^{m=2} \end{bmatrix}.
\label{eq:17}
\end{equation}
Here, the $m=0$ mode corresponds to the case with no progressive phase delay between ports.
In contrast, the $m=1$ and $m=2$ modes exhibit $120^\circ$ (clockwise mode) and $-120^\circ$ (counter-clockwise mode) progressive phase delays, respectively, at the fundamental frequency.
According to (\ref{eq:17}), in order to achieve half the source voltage at port 1 and a null at port 2, the clockwise and counter-clockwise mode voltages must be designed to constructively interfere at port 1 and destructively interfere at port 2.
The required modal voltages are derived using (\ref{eq:17}), and the modulation parameters, $M$ and $\Omega$, are adjusted to ensure that the conditions given by (\ref{eq:16}) are satisfied.
This formulation provides a fast and accurate computational approach that solves the network by analyzing a single unit cell.
It also offers valuable insights into the design of such nonreciprocal devices.

Many time-varying circulators have been proposed in the literature \cite{9257419, 7378329, Sounas2013}.
However, some exhibit narrow bandwidths due to their resonance-based designs \cite{9257419}, while others exhibit high reflection caused by the parametric process \cite{7378329}.
In this work, we applied our analysis to the circulator design reported in \cite{9257419}, optimizing the design parameters to achieve improved bandwidth and reduced insertion loss.
The adjusted design parameters for this circulator are summarized in Table~\ref{tab:4}.
The theoretical analysis uses a quality factor of $Q = 55$, to account for losses in the inductor and varactor.
Fig.~\ref{fig:10_a} illustrates the response of the circulator when no space-time modulation is applied to the loop.
Applying SDTWM to the loop network enhances transmission to port 1, while minimizing reflection at port 0, and isolation at port 2 is achieved.
As noted, only the modulation depth ($M$) and angular modulation frequency ($\Omega$) are adjusted, while all other parameters remain the same as those reported in \cite{9257419}.
The circulator properties compared to those in \cite{9257419} are shown in Table~\ref{tab:5}.
The fractional bandwidth is doubled, and the insertion loss from port 0 to port 1 is improved by approximately 2 dB within the band. 
This improvement is achieved based on the proposed approach, which decomposes the circulator into three driven modal solutions.

\begin{table}
\begin{center}
\centering
\caption{Performance comparison of the designed non-magnetic circulator and the circulator reported in \cite{9257419}.}
\label{tab:5}
\scalebox{1.1}{
\begin{tabular}{| c | c | c | c | c|}
\hline
& $\Omega/2\pi$ & $M$ & Bandwidth & Insertion loss\\
\hline
This work & $2.77$ GHz &$0.35$ & $186$ MHz& $-0.63$ dB\\
\hline
\cite{9257419}& $0.37$ GHz & $0.20$ & $92.8$ MHz & $-2.62$ dB\\
\hline
\end{tabular}}
\end{center}
\end{table}

One advantage of a non-magnetic circulator based on SDTWM loops is that the clockwise and counter-clockwise modes can be swapped by reversing the modulation phase delay between the unit cells.
This causes the signal to constructively interfere at port 2 and destructively interfere at port 1.
It causes the isolation and transmission in Fig.~\ref{fig:3_b} to be swapped.

The simulated S-parameters shown in Fig.~\ref{fig:10} are obtained using the harmonic balance solver in ADS Keysight, incorporating Toshiba 1SV280 varactor diodes.
The diode was accurately characterized using the polynomial Grading Coefficient model reported in \cite{Strohman_2024}, which accounts for device parasitics.
The number of harmonics considered in both the simulations and the semi-analytical framework is $q = 11$.
The simulations using the realistic varactor diode show close agreement with the theoretical results.
{The presented formulation focuses on linear SDTWM networks and therefore does not explicitly model nonlinear device behavior, such as that of varactor diodes, in the theoretical analysis.
Nevertheless, similar Interpath-Relation-based formulations have been extended to nonlinear periodic networks using iterative techniques \cite{Johnson:24}, suggesting that nonlinear effects could be incorporated within a similar framework if required.}

The proposed semi-analytical approach provides an efficient framework for analyzing and designing SDTWM loop networks.
Driven modal solutions are computed and summed to evaluate the network's response to a single source excitation.
This method ensures computational simplicity by focusing on a single unit cell while maintaining high accuracy.

\section{Conclusion}
This paper {presents} a semi-analytical framework for analyzing spatially discrete traveling-wave modulated (SDTWM) loop networks.
By solving the driven modal response using the Interpath Relation, the method enables the analysis of {loop SDTWM networks under phased-array excitation}.
{In addition, this work introduces the analysis of multi-tone modulation in SDTWM loop networks as a new contribution.}
{Moreover, we introduce a formulation for multi-conductor SDTWM networks.}
The analysis framework provides control over {LTI} modes and frequency harmonics while requiring only a single unit cell for computation.
{Finally, the spatial Green’s function formulation previously developed for straight SDTWM networks is extended here to loop configurations} through analytic array scanning.
Application examples, including an electrically small antenna and a non-magnetic circulator, validated the approach and demonstrated its use in design.
These results emphasize the potential of SDTWM loop networks to develop compact and high-performance RF devices (including nonreciprocal devices, parametric amplifiers, and N-path filters) and provide valuable insight into their design and optimization.
The proposed method offers both speed and design insight, making it a powerful tool for the development and optimization of advanced SDTWM electromagnetic systems.
The multi-modal, multi-harmonic analysis enables new degrees of freedom in space-time wave engineering, opening new possibilities for advanced wave manipulation.

\section*{APPENDIX} \label{app}
In this paper, we propose a semi-analytical framework for analyzing spatially discrete traveling-wave modulated (SDTWM) loop networks.
However, we did not explicitly define the construction of the ABCD matrix of a unit cell or its extension to multi-tone modulation.
This appendix provides a detailed formulation, beginning with the analysis of a time-varying capacitor and systematically developing the matrix representation of unit cells used for both single-tone and multi-tone SDTWM networks.
We start by deriving the admittance matrix $\bar{\bar{Y}}$ of a linear periodically time-varying (LPTV) capacitor.
In this process, we introduce and construct the key matrices required for analysis, including the angular frequency matrix $\bar{\bar{\omega}}$ and the Fourier coefficient matrix $\bar{\bar{\mathcal{A}}}_{\mathrm{F}}$.
The formulation is first presented for the single-tone case and is subsequently extended to a general $v$-tone modulation case.

\subsection{Single-Tone Time-Varying Capacitor} \label{app:1}
Consider the LPTV capacitor located in the $n^{\text{th}}$ unit cell, denoted as $C_n(t)$, whose current-voltage relationship satisfies:
\begin{equation}
\mathrm{i}(t) = \frac{d}{dt} \left[ C_n(t) \mathrm{v}(t) \right].
\label{eq:app1}
\end{equation}

Assuming $C_n(t)$ is periodic with period $T = \frac{2\pi}{\Omega}$, where $\Omega$ is the angular frequency of the modulation waveform, the voltage and current signals across the capacitor can be expressed as Fourier series over harmonic indices $\ell \in \mathbb{Z}$:
\begin{equation}
\left\{
\begin{aligned}
\mathrm{v}(t) &= \sum_{\ell} V[\ell]\, e^{j(\omega_0 + \ell \Omega)t} \\
\mathrm{i}(t) &= \sum_{\ell} I[\ell]\, e^{j(\omega_0 + \ell \Omega)t}
\end{aligned}
\right.
\label{eq:app2}
\end{equation}
Here $\omega_0$ is the signal angular frequency.

The LPTV capacitance can similarly be expanded as:
\begin{equation}
C_n(t) = C_0 \sum_{\ell} a[\ell] e^{j \ell \Omega t},
\label{eq:app3}
\end{equation}
where $C_0$ is the average (DC) capacitance value, and $a[\ell] \in \mathbb{C}$ are the normalized Fourier coefficients of the modulation waveform.
The normalization ensures that $a[0] = 1$.

In practice, the infinite sum over harmonics is truncated to a finite range, taken as $-L \leq \ell \leq L$, resulting in $q = 2L + 1$ total frequency harmonics.
This truncation defines the temporal resolution of the analysis and directly determines the dimension of the resulting matrices.  
The choice of $q$ controls the trade-off between accuracy and computational cost.
Increasing $q$ includes higher-order Fourier terms and leads to more accurate results, especially in cases of strong modulation depth or highly non-sinusoidal waveforms.

By substituting \eqref{eq:app2} and \eqref{eq:app3} into \eqref{eq:app1}, and collecting terms at frequency $\omega_0 + \ell \Omega$, we obtain:
\begin{equation}
I[\ell] = j C_0 (\omega_0 + \ell \Omega) \sum_{\ell'=-L}^{L} a[\ell - \ell'] V[\ell'].
\label{eq:app4}
\end{equation}
Here, $\ell$ is the output harmonic index, while $\ell'$ is a dummy summation index used to perform convolution.
This discrete convolution describes harmonic mixing due to time modulation of the capacitor.

The expression can be compactly written in matrix form \cite{Scarborough2022}:
\begin{equation}
\bar{I} = j C_0 \, \bar{\bar{\omega}} \bar{\bar{\mathcal{A}}}_{\mathrm{F}} \bar{V},
\label{eq:app5}
\end{equation}
The $\bar{V}$ and $\bar{I}$ are multi-harmonic column vectors of voltage and current, and can be written as:
\begin{equation}
\bar{V} =
\begin{bmatrix}
V_{-L} \\
\vdots \\
V_{0} \\
\vdots \\
V_{L}
\end{bmatrix}_{q \times 1},
\quad
\bar{I} =
\begin{bmatrix}
I_{-L} \\
\vdots \\
I_{0} \\
\vdots \\
I_{L}
\end{bmatrix}_{q \times 1}.
\label{eq:app6}
\end{equation}

The angular frequency matrix $\bar{\bar{\omega}}$ is a diagonal matrix containing the harmonic frequencies $\omega_0 + \ell \Omega$ along its diagonal:
\begin{equation}
\bar{\bar{\omega}}[\ell,\ell'] = \delta_{\ell - \ell'} \left( \omega_0 + \ell\Omega \right).
\label{eq:app7}
\end{equation}

In explicit matrix form, it appears as:
\begin{equation}
\bar{\bar{\omega}} =
\begin{bmatrix}
\omega_0 - L\Omega & 0 & \cdots & 0 \\
0 & \omega_0 - (L-1)\Omega & \cdots & 0 \\
\vdots & \vdots & \ddots & \vdots \\
0 & 0 & \cdots & \omega_0 + L\Omega
\end{bmatrix}_{q \times q}.
\label{eq:app8}
\end{equation}

The matrix $\bar{\bar{\mathcal{A}}}_{\mathrm{F}}$ captures the harmonic coupling introduced by the time-varying capacitance, and its entries are given by the Fourier coefficients $a[\ell - \ell']$ defined by~\eqref{eq:app4}.
Its elements are:
\begin{equation}
\bar{\bar{\mathcal{A}}}_{\mathrm{F}}[\ell, \ell'] = a[\ell - \ell'].
\label{eq:app9}
\end{equation}
This results in a Toeplitz matrix of the form:
\begin{equation}
\bar{\bar{\mathcal{A}}}_{\mathrm{F}} =
\begin{bmatrix}
a[0] & a[-1] & \cdots & a[-2L] \\
a[1] & a[0] & \cdots & a[-2L+1] \\
\vdots & \vdots & \ddots & \vdots \\
a[2L] & a[2L-1] & \cdots & a[0]
\end{bmatrix}_{q \times q}.
\label{eq:app10}
\end{equation}

\subsubsection{Example: Single-Tone Sinusoidally Modulated Capacitor}
Assume a sinusoidally modulated capacitor:
\begin{equation}
C_n(t) = C_0 \left[ 1 + 2M \cos(\Omega t - n \Omega t_0) \right],
\label{eq:app11}
\end{equation}
where $M$ is the modulation depth, $t_0$ is the temporal modulation delay between capacitors of adjacent unit cells, and $n$ is the unit cell number.
The Fourier series of $C_n(t)$ contains only three nonzero terms:
\begin{equation}
a[0] = 1, \quad
a[\pm 1] = M e^{\mp j n \Omega t_0}, \quad
a[k] = 0 \quad \text{for } |k| > 1.
\label{eq:app12}
\end{equation}

The matrix $\bar{\bar{\mathcal{A}}}_{\mathrm{F}} \in \mathbb{C}^{q \times q}$, constructed from these coefficients, becomes tridiagonal with first-order harmonic coupling:
\begin{equation}
\bar{\bar{\mathcal{A}}}_{\mathrm{F}}[{i,j}] =
\begin{cases}
1, & i = j, \\
M e^{-j n \Omega t_0}, & i = j + 1, \\
M e^{+j n \Omega t_0}, & i = j - 1, \\
0, & \text{otherwise}.
\end{cases}
\label{eq:app13}
\end{equation}

In matrix form, this becomes:
\begin{equation}
\bar{\bar{\mathcal{A}}}_{\mathrm{F}} =
\begin{bmatrix}
1 & M e^{+j n \Omega t_0} & \cdots & 0 \\
M e^{-j n \Omega t_0} & 1 & \cdots & 0 \\
\vdots & \vdots & \ddots & M e^{+j n \Omega t_0} \\
0 & 0 & M e^{-j n \Omega t_0} & 1 \\
\end{bmatrix}_{q \times q}.
\label{eq:app14}
\end{equation}

Now, by constructing the matrices $\bar{\bar{\mathcal{A}}}_{\mathrm{F}}$ for a sinusoidally modulated capacitor and $\bar{\bar{\omega}}$  as defined in~\eqref{eq:app7}, we can define the multi-harmonic admittance matrix of the capacitor using~\eqref{eq:app5} as:
\begin{equation}
\bar{\bar{Y}} = j C_0 \, \bar{\bar{\omega}} \bar{\bar{\mathcal{A}}}_{\mathrm{F}}.
\label{eq:app15}
\end{equation}

Accordingly, the ABCD matrix of the LPTV series capacitor, such as the one shown in Fig.~\ref{fig:1}, is given by:
\begin{equation}
\bar{\bar{T}}_{C} =
\begin{bmatrix}
\bar{\bar{I}} & \bar{\bar{Y}}^{-1} \\
\bar{\bar{0}} & \bar{\bar{I}}
\end{bmatrix}_{2q \times 2q},
\label{eq:app16}
\end{equation}
where $\bar{\bar{I}}$ and $\bar{\bar{0}}$ denote the $q \times q$ identity and zero matrices, respectively.

By writing the ABCD matrix of each element in the unit cell in this form, the total ABDC matrix of the modulated unit cell can be obtained by sequentially multiplying the individual ABCD matrices of its constituent elements.
For time-modulated components, such as the series capacitor considered above, the ABCD matrix contains off-diagonal terms due to harmonic coupling, captured by the Fourier coefficient matrix $\bar{\bar{\mathcal{A}}}_{\mathrm{F}}$.
In contrast, linear time-invariant (LTI) elements, such as static transmission lines or passive inductors, do not couple different harmonics and therefore produce diagonal multi-harmonic impedance or admittance matrices.
Nonetheless, these LTI elements are still represented in the same $q \times q$ harmonic basis, with each diagonal entry corresponding to a different frequency $\omega_0 + \ell \Omega$.
This formulation allows for the systematic construction of the total unit cell response in the frequency-harmonic domain, which can then be used for finding the driven modal solution and spatial Green's function of SDTWM networks.

\subsection{Extension to Multi-Tone Modulation}
As presented in the paper, multi-tone modulation refers to capacitors modulated simultaneously by multiple temporally incommensurate waveforms.
Here, we consider the case where the capacitance is modulated by a sum of $v$ cosine tones, expressed as:
\begin{equation}
C_n(t) = C_0 \left[ 1 + \sum_{\nu=1}^{v} 2M_\nu \cos(\Omega_\nu t - n \Omega_\nu t_\nu) \right],
\label{eq:app17}
\end{equation}
where $M_\nu$ is the modulation depth, $\Omega_\nu$ is the angular frequency of the $\nu^{\text{th}}$ tone, $t_\nu$ is the time delay per unit cell for that tone, and $n$ is the unit cell index.
Here, $\nu$ denotes the index of the modulation tone.

In a multi-tone modulation scheme, the voltage and current across the capacitor exhibit harmonic components at frequencies corresponding to all possible linear combinations of the modulation tones $\Omega_\nu$.
They can be expressed as:
\begin{equation}
\left\{
\begin{aligned}
\mathrm{v}(t) &= \sum_{\ell_1 = -L}^{L} \cdots \sum_{\ell_v = -L}^{L}
V[\ell_1, \dots, \ell_v]\, e^{j\left( \omega_0 + \sum_{\nu=1}^{v} \ell_\nu \Omega_\nu \right)t} \\
\mathrm{i}(t) &= \sum_{\ell_1 = -L}^{L} \cdots \sum_{\ell_v = -L}^{L}
I[\ell_1, \dots, \ell_v]\, e^{j\left( \omega_0 + \sum_{\nu=1}^{v} \ell_\nu \Omega_\nu \right)t}
\end{aligned}
\right.
\label{eq:app18}
\end{equation}
Here the voltage and current expressed as multi-dimensional Fourier series in the presence of $v$ modulation tones with angular frequencies $\Omega_1, \dots, \Omega_v$.
Each index $\ell_\nu$ corresponds to the harmonic index associated with the $\nu^{\text{th}}$ modulation tone, where $\ell_\nu \in \{-L, \dots, L\}$.
Since each modulation tone contributes its own set of harmonics independently, the expression requires $v$ separate summations.
Each index spans $q=2L+1$ values, which leads to a total of $q^v$ spectral components.
This combination of harmonics determines the total number of frequency components, which directly sets the size of the Fourier-expanded voltage and current vectors in the matrix formulation.

The time-varying capacitance can similarly be expanded in a multi-dimensional Fourier series as:
\begin{equation}
C_n(t) = C_0 \sum_{\ell_1 = -L}^{L} \cdots \sum_{\ell_v = -L}^{L}
a[\ell_1, \dots, \ell_v] \, e^{j \left( \sum_{\nu=1}^{v} \ell_\nu \Omega_\nu \right) t},
\label{eq:app19}
\end{equation}
where $a[\ell_1, \dots, \ell_v]$ are the Fourier coefficients of the $v$-tone waveform.

Substituting \eqref{eq:app18} and \eqref{eq:app19} into the current-voltage relationship given in \eqref{eq:app1}, we obtain:
\begin{equation}
\begin{aligned}
&I[\ell_1, \dots, \ell_v] =\,  j C_0 \left( \omega_0 + \sum_{\nu=1}^{v} \ell_\nu \Omega_\nu \right) \\
& \times \sum_{\ell_1' = -L}^{L} \cdots \sum_{\ell_v' = -L}^{L}
a[\ell_1 - \ell_1', \dots, \ell_v - \ell_v'] \, V[\ell_1', \dots, \ell_v'].
\end{aligned}
\label{eq:app20}
\end{equation}
This equation generalizes the single-tone convolution to a multi-dimensional convolution, where each modulation tone introduces coupling between neighboring harmonic indices in its respective dimension.

To express \eqref{eq:app20} in a compact matrix form (similar to the single-tone case in \eqref{eq:app5}) we first need to define the angular frequency matrix $\bar{\bar{\omega}}$ and the Fourier coefficient matrix $\bar{\bar{\mathcal{A}}}_{\mathrm{F}}$. 
As described earlier, since the modulation waveform consists of $v$ distinct tones, the resulting matrices in \eqref{eq:app20} are naturally $v$-dimensional.
However, to remain consistent with our overall formulation (particularly the Interpath Relation and the driven modal solution) both the ABCD matrix and the delay matrices must remain two-dimensional.
Therefore, a mapping procedure is required to convert the $v$-dimensional matrix into a two-dimensional matrix representation.

\subsubsection{Mapping procedure for $v$-Tone Modulation}

To describe the frequency space arising from $v$-tone modulation, we define an auxiliary matrix called the multi-index harmonic matrix, denoted by $\mathcal{H} \in \mathbb{Z}^{q^v \times v}$.
This auxiliary matrix provides a systematic and structured way to enumerate all combinations of harmonic indices resulting from modulation by $v$ independent tones.
The multi-index harmonic matrix is defined as:
\begin{equation}
\mathcal{H} =
\begin{bmatrix}
\ell_1[1] & \ell_2[1] & \cdots & \ell_v[1] \\
\ell_1[2] & \ell_2[1] & \cdots & \ell_v[1] \\
\vdots    & \vdots    & \ddots & \vdots    \\
\ell_1[q] & \ell_2[1] & \cdots & \ell_v[1] \\
\ell_1[1] & \ell_2[2] & \cdots & \ell_v[1] \\
\vdots    & \vdots    & \ddots & \vdots    \\
\ell_1[q] & \ell_2[2] & \cdots & \ell_v[1] \\
\vdots    & \vdots    & \ddots & \vdots    \\
\ell_1[1] & \ell_2[q] & \cdots & \ell_v[q] \\
\vdots    & \vdots    & \ddots & \vdots    \\
\ell_1[q] & \ell_2[q] & \cdots & \ell_v[q]
\end{bmatrix}_{q^v \times v}
\label{eq:app21}
\end{equation}
Each column of $\mathcal{H}$ corresponds to the $\nu^{\text{th}}$ modulation tone, where $\nu \in \{1, \dots, v\}$, resulting in a total of $v$ columns.
Specifically, the $\nu^{\text{th}}$ column contains the harmonic index $\ell_\nu \in \{-L, \dots, L\}$ associated with the $\nu^{\text{th}}$ tone.
Each row represents a unique multi-index harmonic combination.
As an example, $\ell_3[2]$ refers to the second entry corresponding to the harmonic index of the third modulation tone.
Assuming the index set $\{-L, \dots, L\}$, the value of $\ell_3[2]$ is $-L + 1$.
Since each tone contains $q$ harmonics, the total number of distinct combinations is $q^v$.
The matrix $\mathcal{H}$ is constructed in descending frequency order, where $\ell_1$ varies fastest and $\ell_v$ varies slowest.

In the case where $v = 2$ modulation tones and each tone includes $q = 3$ harmonics (i.e., $\ell_1, \ell_2 \in \{-1, 0, 1\}$), the multi-index harmonic matrix becomes $\mathcal{H} \in \mathbb{Z}^{9 \times 2}$, as follows,
\begin{equation}
\mathcal{H} =
\begin{bmatrix}
-1 & -1 \\
\hspace{0.7em}0 & -1 \\
\hspace{0.7em}1 & -1 \\
-1 & \hspace{0.7em}0 \\
\hspace{0.7em}0 & \hspace{0.7em}0 \\
\hspace{0.7em}1 & \hspace{0.7em}0 \\
-1 & \hspace{0.7em}1 \\
\hspace{0.7em}0 & \hspace{0.7em}1 \\
\hspace{0.7em}1 & \hspace{0.7em}1
\end{bmatrix}.
\label{eq:app22}
\end{equation}
This matrix provides a complete enumeration of the harmonic index space for the two-tone case (with $q=3$ number of harmonics for each tone).
The physical interpretation of the multi-index harmonic matrix is that it defines the set of frequency components present in the network due to modulation.
For example, in the two-tone case given by~\eqref{eq:app22}, the third row corresponds to the angular frequency $\omega = \omega_0 + \Omega_1 - \Omega_2$.

The multi-index harmonic matrix $\mathcal{H}$ can now be used to map the $v$-dimensional frequency expression $\left( \omega_0 + \sum_{\nu=1}^{v} \ell_\nu \Omega_\nu \right)$ in~\eqref{eq:app20} into a diagonal two-dimensional angular frequency matrix.
The resulting diagonal matrix $\bar{\bar{\omega}}$ is of size $q^v \times q^v$, where the $i^{\text{th}}$ diagonal entry is defined as:
\begin{equation}
\bar{\bar{\omega}}[i,i] = \omega_0 + \sum_{\nu=1}^{v} \mathcal{H}[{i,\nu}] \Omega_\nu.
\label{eq:app23}
\end{equation} 

\begin{figure}
\centering
\includegraphics[width=85mm]{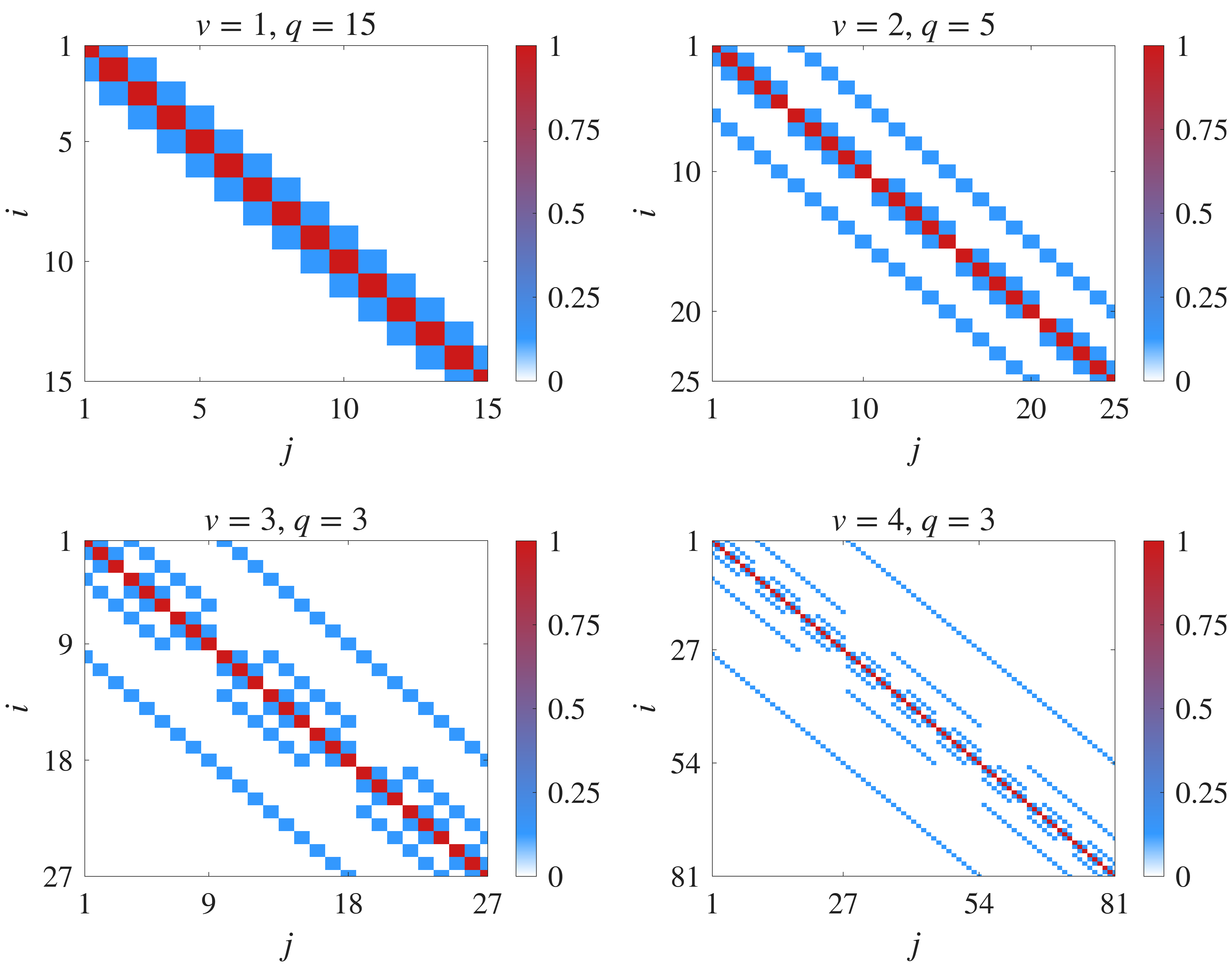}
    \caption{Fourier coefficient matrix $\bar{\bar{\mathcal{A}}}_{\mathrm{F}}$ for various values of modulation tones $v$ and harmonics $q$, constructed from the multi-tone modulation waveform defined in~\eqref{eq:app17}. All cases assume modulation depth \( M_\nu = 0.125 \) and phase delay $\Omega_\nu t_\nu = 0$ for all tones.}
\label{fig:12}
\end{figure}

This mapping enables all matrices, including those corresponding to LTI components such as transmission lines, to be represented in the same harmonic basis.
As a result, it allows multi-tone modulation in SDTWM networks to be analyzed using the semi-analytical framework presented in Section~\ref{sec:3_1}.

Now that the angular frequency matrix has been defined, we proceed to construct the Fourier coefficient matrix associated with the $v$-tone modulation waveform.
Similar to the single-tone case, the matrix $\bar{\bar{\mathcal{A}}}_{\mathrm{F}}$ captures the harmonic coupling introduced by the time-varying capacitance.
Its entries are determined by the multi-dimensional Fourier coefficients, given by:
\begin{equation}
\bar{\bar{\mathcal{A}}}_{\mathrm{F}}[\ell_1, \ell_1', \dots, \ell_v, \ell_v'] = a[\ell_1 - \ell_1', \dots, \ell_v - \ell_v'].
\label{eq:app24}
\end{equation}
As a result, the Fourier coefficient matrix is a $2v$-dimensional tensor, i.e., $\bar{\bar{\mathcal{A}}}_{\mathrm{F}} \in \mathbb{C}^{q \times q \times \cdots \times q}$, with $2v$ dimensions in total.

With the help of the multi-index harmonic matrix $\mathcal{H}$, we can compute the harmonic index differences for each modulation tone.
These differences determine the appropriate order of the Fourier coefficient to be used for each tone.
The harmonic index difference corresponding to the $\nu^{\text{th}}$ modulation tone can be defined as:
\begin{equation}
\Delta_\nu^{(i,j)} = \mathcal{H}[i,\nu] - \mathcal{H}[j,\nu],
\label{eq:app25}
\end{equation}
where $\mathcal{H}[i,\nu]$ and $\mathcal{H}[j,\nu]$ are the entries in the $\nu^{\text{th}}$ column of $\mathcal{H}$, taken from the $i^{\text{th}}$ and $j^{\text{th}}$ rows, respectively.

For modulation wave form given in \eqref{eq:app17}, the $\bar{\bar{\mathcal{A}}}_{\mathrm{F}}$ is then given by:
\begin{equation}
\bar{\bar{\mathcal{A}}}_{\mathrm{F}}[{i,j}] = \prod_{\nu=1}^{v}
\begin{cases}
1, & \Delta_\nu^{(i,j)} = 0 \\
M_\nu e^{-j n \Omega_\nu t_\nu}, & \Delta_\nu^{(i,j)} = 1 \\
M_\nu e^{+j n \Omega_\nu t_\nu}, & \Delta_\nu^{(i,j)} = -1 \\
0, & \text{otherwise}
\end{cases}.
\label{eq:app26}
\end{equation}
This mapping ensures that $\bar{\bar{\mathcal{A}}}_{\mathrm{F}} \in \mathbb{C}^{q^v \times q^v}$ and that the indexing is consistent with the mapping described for the angular frequency matrix.

Fig.~\ref{fig:12} shows the Fourier coefficient matrix $\bar{\bar{\mathcal{A}}}_{\mathrm{F}}$ for different numbers of modulation tones ($v$) and harmonics ($q$), based on the waveform given in~\eqref{eq:app17}.
In the single-tone sinusoidal modulation case, $\bar{\bar{\mathcal{A}}}_{\mathrm{F}}$ is a tridiagonal matrix with unit values along the main diagonal.
As the number of modulation tones increases, the matrix is no longer tridiagonal due to the specific mapping procedure used to construct the 2D representation.
It can be observed that for each case, the resulting matrix has dimensions $q^v \times q^v$.

The $v$-tone admittance of the LPTV capacitance, given in~\eqref{eq:app17}, can be obtained by substituting~\eqref{eq:app23} and~\eqref{eq:app26} into~\eqref{eq:app15}.
The only difference compared to the single-tone case is the size of the matrices, which increases with the number of modulation tones.
This mapping procedure allows the proposed analysis to be applied in a straightforward manner to multi-tone SDTWM networks.


\bibliographystyle{IEEEtran}

\bibliography{references_abbreviated}

\end{document}